\documentclass[aps,prl,superscriptaddress,twocolumn,longbibliography]{revtex4-2}
\usepackage{units}
\usepackage{amsmath}
\usepackage{amsthm}
\usepackage{amssymb}
\usepackage{graphicx}
\usepackage{color}
\usepackage{xcolor}
\usepackage{bbold}
\usepackage{appendix}

\usepackage[colorlinks=true, urlcolor=blue, citecolor=blue,linkcolor=blue,citebordercolor={1 0 0},linkbordercolor={0 0 1}]{hyperref}

\newtheorem{theorem}{Theorem}

\providecommand{\definitionname}{Definition}

\usepackage[linesnumbered, ruled,vlined]{algorithm2e}

\graphicspath{{./images/},{./imagesAppendix/}}

\renewcommand{\eqref}[1]{Eq.~(\ref{#1})} 

\newtheorem{thm}{\protect\theoremname}

\ifx\proof\undefined
\newenvironment{proof}[1][\protect\proofname]{\par
	\normalfont\topsep6\p@\@plus6\p@\relax
	\trivlist
	\itemindent\parindent
	\item[\hskip\labelsep\scshape #1]\ignorespaces
}{%
	\endtrivlist\@endpefalse
}
\providecommand{\proofname}{Proof}
\fi

\newcommand{\bra}[1]{\langle #1|}
\newcommand{\ket}[1]{|#1 \rangle}
\makeatother
\newcommand{\braket}[2]{\langle #1 \vert #2 \rangle}
\newcommand{\abs}[1]{\left|#1\right|}
\newcommand{\idg}[1]{{\bfseries #1)}}

\newcommand\numberthis{\addtocounter{equation}{1}\tag{\theequation}}
\providecommand{\factname}{Fact}
\providecommand{\theoremname}{Theorem}
\providecommand{\claimname}{Claim}
\providecommand{\lemmaname}{Lemma}
\providecommand{\definitionname}{Definition}
\providecommand{\observationname}{Observation}

\newtheorem{observation}{\protect\observationname}

\providecommand{\propositionname}{Proposition}

\newtheorem{defn}[thm]{\protect\definitionname}

\definecolor{KB}{rgb}{0.4,0.3,0.9}

\definecolor{THc}{rgb}{0.9,0.3,0.2}

\newcommand{\revA}[1]{{#1}}

\newcommand{\revB}[1]{#1}

\newcommand{\prlsection}[1]{{\em {#1}.---~}}

\newcommand{\subfigimg}[3][,]{%
	\setbox1=\hbox{\includegraphics[#1]{#3}}
	\leavevmode\rlap{\usebox1}
	\rlap{\hspace*{2pt}\raisebox{\dimexpr\ht1-0.5\baselineskip}{{\bfseries \large\textsf{#2}}}}
	\phantom{\usebox1}
}

\newcommand{\sectionMain}[1]{
\let\oldaddcontentsline\addcontentsline
\renewcommand{\addcontentsline}[3]{}
\section{#1}
\let\addcontentsline\oldaddcontentsline
}

\allowdisplaybreaks

\begin{document}

\title{Generalization of Quantum Machine Learning Models Using Quantum Fisher Information Metric}

\author{Tobias Haug}
\email{tobias.haug@u.nus.edu}
\affiliation{Quantum Research Center, Technology Innovation Institute, Abu Dhabi, UAE}
\affiliation{Blackett Laboratory, Imperial College London SW7 2AZ, UK}

\author{M.~S. Kim}
\affiliation{Blackett Laboratory, Imperial College London SW7 2AZ, UK}

\begin{abstract}
Generalization is the ability of machine learning models to make accurate predictions on new data by learning from training data. 
However, understanding generalization of quantum machine learning models has been a major challenge.
Here, we introduce the data quantum Fisher information metric (DQFIM). It describes the capacity of variational quantum algorithms depending on variational ansatz, training data and their symmetries. 
We apply the DQFIM to quantify circuit parameters and training data needed to successfully train and generalize. 
Using the dynamical Lie algebra, we explain how to generalize using a low number of training states. 
Counter-intuitively, breaking symmetries of the training data can help to improve generalization. 
Finally, we find that out-of-distribution generalization, where training and testing data are drawn from different data
distributions, can be better than using the same distribution.
Our work provides a useful framework to explore the power of quantum machine learning models.
\end{abstract}

\maketitle


The key challenge in quantum machine learning  is to design models that can learn from data and apply their acquired knowledge to perform well on new data~\cite{biamonte2017quantum}. This latter ability is called generalization and has been intensely studied recently~\cite{poland2020no,caro2020pseudo,abbas2020power,abbas2021effective,sharma2022reformulation,banchi2021generalization,caro2022out,peters2022generalization,gibbs2022dynamical,volkoff2021universal,cai2022sample,popescu2021learning,caro2021encoding,bu2022statistical,bowles2023contextuality,du2022demystify,bu2023effects,gili2022evaluating}. Constructing models that generalize well is essential for quantum machine learning tasks such as learning unitaries~\cite{bisio2010optimal,marvian2016universal,bharti2021noisy,cerezo2020variational,xue2022variational,gebhart2023learning,kiani2020learning,yu2023optimal}, classification~\cite{farhi2018classification,schuld2020circuit}, compiling~\cite{khatri2019quantum,ezzell2022quantum,volkoff2021universal}, generative modeling~\cite{coyle2020born,gili2023quantum}, quantum simulation~\cite{cirstoiu2020variational,gibbs2022dynamical,gibbs2022long}, quantum autoencoders~\cite{romero2017quantum,zhang2022resource} and black-hole recovery protocols~\cite{leone2022retrieving}. 
\revA{However, the conditions for generalization are not well understood. Recently proposed uniform generalization bounds~\cite{caro2021generalization,banchi2021generalization} have been shown to be loose~\cite{gil2023understanding}, do not account for symmetries and are unable to explain numerical observations of generalization with few training data~\cite{caro2021generalization,gibbs2022dynamical,gil2023understanding}.}

Thus, there is an urgent need for a framework to understand the conditions for successful training and generalization~\cite{schatzki2022theoretical,ragone2022representation,nguyen2022theory,zheng2021speeding,meyer2022exploiting,larocca2022group,sauvage2022building,skolik2022equivariant,haug2021capacity,bu2021rademacher,anschuetz2022efficient,caro2022out} to potentially gain advantage over classical models~\cite{huang2021power,liu2020rigorous,huang2021provably,liu2023towards}. 
In classical machine learning, generalization has been evaluated using the classical Fisher information~\cite{bartlett2017spectrally,jiang2019fantastic,liang2019fisher,abbas2020power,abbas2021effective}. 
Recent works proposed the quantum Fisher information metric (QFIM) to characterize capacity and overparameterization of parameterized quantum states~\cite{haug2021capacity,larocca2021theory,meyer2021fisher,garcia2023effects}, however a connection with generalization has not been established.

Here, we introduce the data quantum Fisher information metric (DQFIM) to study generalization and overparameterization. 
In contrast to the QFIM, the DQFIM correctly captures the effect of data and circuit symmetries on the capacity of quantum machine learning models.
The rank of the DQFIM quantifies the circuit depth and amount of data needed for generalization and convergence to a global minimum of the cost function. 
\revA{We apply our methods to learning unitaries, quantum control, generative models, finding excited states and classification tasks.}
\revA{Using the connection between DQFIM and dynamical Lie algebra (DLA), we explain why quantum machine learning can generalize with few training data.} 
While symmetries have been known to benefit quantum machine learning, we surprisingly find that symmetries in data can also hinder generalization. 
Finally, we show that out-of-distribution generalization, i.e. the training data is drawn from a different distribution than the test data, can exhibit better performance compared to in-distribution generalization.
Our methods provide a quantum geometric picture to understand generalization which guides the design of better quantum machine learning models.

\noindent\prlsection{Model}
\revA{We consider a unitary $U(\boldsymbol{\theta})$ parameterized by $M$-dimensional parameter vector~$\boldsymbol{\theta}$ and training set $S_L=\{\ket{\psi_\ell}, O_\ell\}_{\ell=1}^L$ of size $L$. $S_L$ consists of input states $\ket{\psi_\ell}$ drawn from a distribution $\ket{\psi_\ell}\in W$, as well as hermitian operator $O_\ell$ which represents the label~\cite{caro2021generalization,gibbs2022dynamical,gibbs2022long}. 
We now learn by minimizing the cost function
\begin{equation}\label{eq:cost_train}
C_\text{train}(\boldsymbol{\theta},S_L)=1-\frac{1}{L}\sum_{\ell=1}^L\bra{\psi_\ell}U(\boldsymbol{\theta})^\dagger O_\ell U(\boldsymbol{\theta})\ket{\psi_\ell}\,.
\end{equation}
Here, we assume without loosing generality that the eigenvalues of $O_\ell$ are (tightly) upper bounded by $1$ such that $C_\text{train}\ge0$. 
The trained model generalizes when the test error in respect to unseen test data $\ket{\psi}\in W$ and corresponding label $O_{\ket{\psi}}$ is small
\begin{equation}\label{eq:cost_test}
C_\text{test}(\boldsymbol{\theta},W)=1- \underset{\ket{\psi}\in W}{\mathbb{E}}[\bra{\psi}U(\boldsymbol{\theta})^\dagger O_{\ket{\psi}} U(\boldsymbol{\theta})\ket{\psi}]\,.
\end{equation}
Let us now give two important examples of our model: First, unitary learning or quantum compiling aims to represent a target unitary $V$  with a parameterized unitary $U(\boldsymbol{\theta})$ such that $V\ket{\psi_\ell}=U(\boldsymbol{\theta})\ket{\psi_\ell}$~\cite{khatri2019quantum,kiani2020learning}. Here, $\ket{\psi_\ell}$ are initial states with corresponding label operator $O_\ell=V\ket{\psi_\ell}\bra{\psi_\ell}V^\dagger$ being the target state to be learned. 
This learning model also describes quantum control problems~\cite{chakrabarti2007quantum}. Further, unsupervised generative models to learn a probability distribution $p(x)$ can be converted into unitary learning tasks~\cite{rudolph2023trainability} by encoding empirical distribution $q(x)$ into a state $\ket{\Phi}$ with $q(x)\sim\vert\braket{x}{\Phi}\vert^2$, and choosing $O_\ell=\ket{\Phi}\bra{\Phi}$. 

Another important task is classification~\cite{farhi2018classification}: Here, the goal is to identify two classes, e.g. images of cats and dogs. One encodes the feature vector $\boldsymbol{x}_\ell$ into $\ket{\psi_\ell(\boldsymbol{x}_\ell)}$ with corresponding label $y_\ell=\pm1$ and label operator $O_\ell=y_\ell\sigma^z$, where cats have $y=1$ and dogs $y=-1$. The trained model infers the class $y$ by measuring $y\sim\bra{\psi}U(\boldsymbol{\theta})^\dagger\sigma^z U(\boldsymbol{\theta})\ket{\psi}$. \revB{We note that data-reuploading~\cite{perez2020data}, where data and parameterized unitary are interlayered, can also be mapped onto this model~\cite{jerbi2023quantum}.}}

The parameterized unitary $U(\boldsymbol{\theta})=\prod_{k=1}^G U^{(k)}(\boldsymbol{\theta}_k)$ commonly consists of $G$ repeating layers of unitaries $U^{(k)}(\boldsymbol{\theta}_k)=\prod_{n=1}^K\exp(-i\theta_{kn}H_n)$, where $H_n$ are hermitian operators, $\boldsymbol{\theta}_k$ a $K$-dimensional vector, and $\boldsymbol{\theta}=\{\boldsymbol{\theta}_1,\dots,\boldsymbol{\theta}_G\}$ the $M=GK$ dimensional parameter vector~\cite{bharti2021noisy,chakrabarti2007quantum}. 
The optimization program starts with a randomly chosen $\boldsymbol{\theta}$ and iteratively minimizes~\eqref{eq:cost_train} with the gradient $\nabla C_\text{train}(\boldsymbol{\theta})$, which can be efficiently estimated by a quantum computer~\cite{mitarai2018quantum}.
Gradient descent iteratively updates $\boldsymbol{\theta}\rightarrow \boldsymbol{\theta}-\alpha\nabla C_\text{train}$ with some $\alpha$ until reaching a minimum after $E$ training steps, where $\nabla C_\text{train}(\boldsymbol{\theta}^*)=0$ with converged parameter $\boldsymbol{\theta}^*$. 
We assume that ansatz $U(\boldsymbol{\theta})$ can solve the learning task, i.e. we ensure that there is a parameter $\boldsymbol{\theta}_\text{g}$ such that $C_\text{test}(\boldsymbol{\theta}_\text{g},W)=C_\text{train}(\boldsymbol{\theta}_\text{g},S_L)=0$.

After training we have three possible outcomes:
$(i)$ Become stuck in local minimum $C_\text{test}\geq C_\text{train}\gg0$;
$(ii)$ Reach global minimum $C_\text{train}\approx 0$, however no generalization with $C_\text{test}\gg0$.
$(iii)$ Generalization with $C_\text{train}\approx C_\text{test}\approx0$.
In the following, we show that the DQFIM determines the critical number of circuit parameters $M_\text{c}(L)$ for overparameterization as function of $L$ and training states $L_\text{c}$ for generalization.

\begin{figure}[htbp]
	\centering	
\subfigimg[width=0.48\textwidth]{}{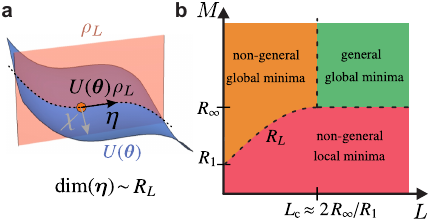}\hfill
	\caption{ 
 \idg{a}~Ansatz unitary $U(\boldsymbol{\theta})$ and $M$-dimensional parameter vector $\boldsymbol{\theta}$ is optimized in respect to cost function~\eqref{eq:cost_train} using $L$ training data described by datastate $\rho_L$ (\eqref{eq:datastate}). 
 Only the subspace of the unitary that acts on the training data $U_L\equiv U(\boldsymbol{\theta})\rho_L$ can be learned. Its learnable degrees of freedom are given by the maximal rank of the data quantum Fisher information metric (DQFIM) $R_L$. 
 \idg{b}~Phase diagram of generalization with $M$ and $L$. 
 Convergence to global minimum ($C_\text{train}\approx0)$ is likely for overparameterization $M\ge R_L$. Generalization to unseen test data ($C_\text{test}\approx0$) for overcomplete training data when $L\ge L_\text{c}\approx 2R_\infty/R_1$ and $M\ge R_\infty$. 
	}
	\label{fig:phases}
\end{figure}

\noindent\prlsection{DQFIM} 
First, we define what can be learned about ansatz unitary $U(\boldsymbol{\theta})$ via training set $S_L$: 
\begin{defn}[\revB{Unitary mapped onto data}]\label{def:dstate}
The data state for training set $S_L=\{\ket{\psi_{\ell}},O_\ell\}_{\ell=1}^L$ of $L$ states is
\begin{equation}\label{eq:datastate}
\rho_L=\frac{1}{L}\sum_{\ell=1}^{L}\ket{\psi_\ell}\bra{\psi_\ell}
\end{equation}
Training with cost function~\eqref{eq:cost_train} and $S_L$ gives only information about the unitary \revB{mapped} onto the subspace of the training data 
$U_L(\boldsymbol{\theta})\sim U(\boldsymbol{\theta})\rho_L$.
\end{defn}
To understand Def.~\ref{def:dstate}, consider the $d$-dimensional unitary $U\equiv U(\boldsymbol{u})= \sum_{n,k=1}^d u_{nk}\ket{n}\bra{k}$ with complex parameters $\boldsymbol{u}=\{u_{11},u_{12},\dots,u_{dd}\}$ and training data $\{\ket{\ell}\}_{\ell=1}^L$, where $\ket{\ell}\in W$ are computational basis states and our goal is to learn some unitary  $V=U(\boldsymbol{u}^\ast)$.
For $L=1$, training with~\eqref{eq:cost_train} optimizes $U\ket{1}=\sum_{n=1}^d u_{n1}\ket{n}$. Thus, only the column vector $u_1=(u_{11},u_{21},\dots,u_{d1})$ of $U$ can be trained, while $u_{n>1}$ are not learnable. 
For arbitrary $L$, applying $U$ on the training states gives us $\{U\ket{\ell}=\sum_{n=1}^d u_{n\ell}\ket{n}\}_{\ell=1}^L$. The learnable parameters of $U$ correspond to
the $d\times L$-dimensional (unnormalized) isometry $U_L=(u_1,\dots, u_L)\equiv U\rho_L$ with (unnormalized) projector $\rho_L=L^{-1}\sum_{\ell=1}^L\ket{\ell}\bra{\ell}$ (see Fig.~\ref{fig:phases}a).
Even if we find a global minima with $C_\text{train}=0$, for $L<d$ we gain no information about the column vectors $(u_{L+1},\dots,u_d)$. The trained model $U(\boldsymbol{u}^*)$ randomly guesses these column vectors, resulting in a large generalization error $C_\text{test}$. Only for $L=d$, we have a complete training set that can achieve generalization with $C_\text{test}=0$.

To understand generalization, we count the independent parameters of $U_L$, which we call the effective dimension $D_L$. For $L=1$, $U\ket{1}=\sum_{n=1}^d u_{n1}\ket{n}=\sum_{n=1}^d (a_{n1}+ib_{n1})\ket{n}$ has $2d$ real parameters $a_{n1}$, $b_{n1}$. However, due to global phase and norm, there are only $D_1=2d-2$ independent parameters. For $L=d$, parameterizing a complete unitary $U$ requires $D_d=d^2-1$ parameters. 
For example, a single qubit has $D_1=2$ (Bloch sphere) and $D_2=3$ (arbitrary unitary) free parameters~\cite{nielsen2002quantum}, and thus we require $L\ge2$ states to generalize.
However, depending on ansatz and data structure $D_L$ can decrease. 
Let us consider $U(\boldsymbol{\theta})=e^{-i\sigma_z \theta_k}\dots e^{-i\sigma_z \theta_M}$ and distribution $W=\{\ket{+},\ket{-}\}$ with $\ket{\pm}\sim\ket{0}\pm\ket{1}$ and $z$-Pauli $\sigma_z$. While we have $M$ parameters, the generators commute and $L=1$ is sufficient to generalize as $D_1=D_d=1$. In contrast, for $W=\{\ket{0},\ket{1}\}$ we have $D_L=0$ as only the trivial global phase is rotated.

\begin{figure*}[htbp]
	\centering	
 \includegraphics[width=0.99\textwidth]{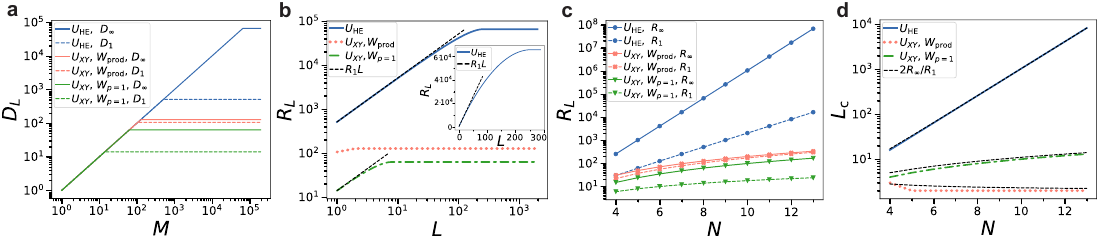}
\caption{DQFIM for different unitaries $U$ with $M$ parameters and $L$ training states. As defined in SM~\ref{sec:ansatz}, we show hardware efficient circuit $U_\text{HE}$ with no symmetries and Haar random training states (blue curves), as well as $U_{XY}$ with particle number symmetry using as training data either product states $W_\text{prod}$ (orange) or symmetry-conserving states  $W_{p=1}$ (green). 
 \idg{a}~Effective dimension $D_L$ increases linearly with $M$, until it reaches a maximal value $R_L$ for $M\ge M_\text{c}(L)$. We have $N=8$ qubits.
 \idg{b}~$R_L$ increases with $L$ until converging to $R_\infty$ for $L\ge L_\text{c}$. Black dashed line shows approximation $R_L\sim R_1L$. Inset shows generic ansatz without log-plot, highlighting the non-linear behavior of $R_L$.
 \idg{c}~Scaling of $R_1$ and $R_\infty$ with qubit number $N$. 
 \idg{d}~Number $L_\text{c}$ of training states needed for generalization. Black dashed line shows $L_\text{c}\approx 2R_\infty/R_1$.
	}
	\label{fig:rank}
\end{figure*}

We now propose the DQFIM to quantify the effective dimension (see Supplemental material (SM)~\ref{sec:dQFIM}):
\begin{defn}[DQFIM]\label{def:dQFIM}
For unitary $U(\boldsymbol{\theta})\equiv U$ and training set $S_L$, the DQFIM is defined as 
\begin{align*}
&\mathcal{Q}_{nm}(\rho_L,U)=\numberthis\label{eq:dQFIM}\\
&4\mathrm{Re}(\mathrm{tr}(\partial_n U \rho_L \partial_m U^\dagger)-\mathrm{tr}(\partial_n U\rho_L U^\dagger)\mathrm{tr}(U \rho_L\partial_m U^\dagger))
\end{align*}
where $\partial_n$ is the derivative in respect to the $n$th entry of the $M$-dimensional vector $\boldsymbol{\theta}=(\theta_1,\dots,\theta_M)$. 
\end{defn}
\revA{In SM~\ref{sec:dQFIM}, we derive $\mathcal{Q}(\rho_L,U(\boldsymbol{\theta}))$ as the metric that describes how a variation $\boldsymbol{\theta}\rightarrow \boldsymbol{\theta}+\text{d}\boldsymbol{\theta}$ changes the mapping of $U(\boldsymbol{\theta})$ onto the span of $\rho_L$, and relate the DQFIM to the QFIM of the purification of $\rho_L$. }
For $L=1$, we recover the QFIM $\mathcal{F}_{nm}=4\text{Re}(\braket{\partial_n\psi}{\partial_m\psi}-\braket{\partial_n\psi}{\psi}\braket{\psi}{\partial_m\psi})$~\cite{liu2020quantum,meyer2021fisher}.

The rank of $\mathcal{Q}$ gives the effective dimension
\begin{equation}
D_L(\rho_L,U(\boldsymbol{\theta}))=\mathrm{rank}(\mathcal{Q}(\rho_L,U(\boldsymbol{\theta}))\le M\,.
\end{equation}
The case $L=1$ has been studied previously~\cite{haug2021capacity}: The effective dimension $D_1$ increases with $M$, until reaching a maximal value $R_1$ (see Fig.~\ref{fig:rank}a). Once maximal, the parameterized state $U(\boldsymbol{\theta})\ket{\psi_1}$ is overparameterized as it can explore all its degrees of freedom~\cite{larocca2021theory}. \revB{While $D_L(\boldsymbol{\theta})$ depends on $\boldsymbol{\theta}$, it turns out that due to self-averaging, a randomly chosen $\boldsymbol{\theta}_\text{rand}$ nearly always assumes its maximal rank $\text{max}_{\boldsymbol{\theta}} D_L(\boldsymbol{\theta})\approx D_L(\boldsymbol{\theta}_\text{rand})$~\cite{haug2021capacity}.}
Just as $D_1$, our $D_L$ increases with $M$ until a maximal $R_L$, which describes the maximal number of degrees of freedom that $U_L$ can explore and heralds overparameterization for arbitrary $L$:
\begin{defn}[Overparameterization]\label{def:ovp_U}
Ansatz $U(\boldsymbol{\theta})$ with training data $\rho_L$ is overparameterized when effective dimension $D_L$ does not increase further upon increasing the number of parameters $M$. The maximal rank $R_L$ reached at critical number of parameters $M\ge M_\text{c}(L)$:
\begin{equation}\label{eq:RLMc}
R_L\equiv\max_{M\ge  M_\text{c}(L),\boldsymbol{\theta}}D_L(\rho_L,U(\boldsymbol{\theta})) \,.
\end{equation}
\end{defn}
For overparameterization with $M\ge M_\text{c}(L)$, a variation of $\boldsymbol{\theta}$ can explore all degrees of freedom of $U_L$ and thus likely find the global minimum~\cite{rabitz2004quantum,bukov2018reinforcement,anschuetz2021critical,you2022convergence,larocca2021theory}:

\begin{figure*}[htbp]
	\centering	
\includegraphics[width=0.99\textwidth]{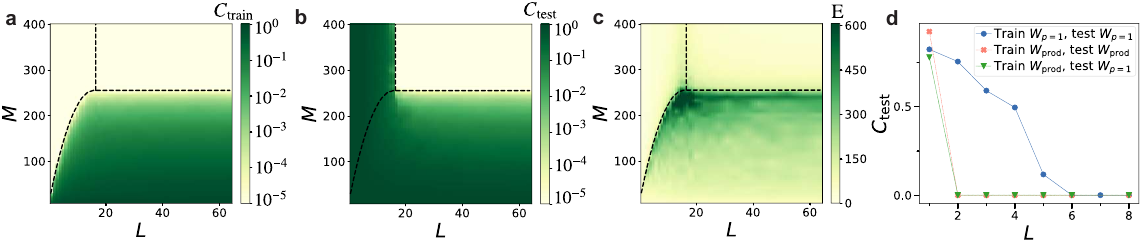}
\caption{\idg{a} Median $C_\text{train}$ against $M$ and $L$ for learning unitaries.  Dashed black lines indicate $M_\text{c}(L)=R_L$ and $L_\text{c}=2R_\infty/R_1$. We have a $N=4$ qubit hardware-efficient ansatz  trained with random training states and gradient descent~\cite{fletcher2013practical} simulated with~\cite{johansson2012qutip}. Target unitary is $V=U(\boldsymbol{\theta}_\text{g})$ with random parameter  $\boldsymbol{\theta}_\text{g}$, where we take median over 10 random instances.
 \idg{b} Average $C_\text{test}$ against $M$ and $L$.
 \idg{c} Number of training steps $E$ until reaching $C_\text{train}< 10^{-4}$.
 \idg{d} $C_\text{test}$ against $L$ with particle-number conserving $U_{XY}$ ansatz for $N=6$ qubits and $M=90$. We train and test with product states $W_{\mathrm{prod}}$ and particle-number conserving states $W_{p=1}$.
}
	\label{fig:test}
\end{figure*}

\begin{observation}[Convergence to global minimum]\label{obs:overparam}
Global minimum $C_\text{train}(\boldsymbol{\theta}^*)\approx0$ with training set $S_L$ is reached with high probability when $M\ge M_\text{c}(L)\ge R_L$.
\end{observation}
As seen in Fig.~\ref{fig:rank}b, $R_L$ increases with $L$, where the growth slows down due to unitary constraints. We find the tight upper bound (SM~\ref{sec:RLexact} or~\cite{polcari2016representing})
\begin{equation}\label{eq:DLsin}
    R_L \le  2dL-L^2-1  \,\,\mathrm{for}\, L \le d\,;\,\, R_L \le d^2-1  \,\,\mathrm{for }\,\, L > d\,.
\end{equation}
$R_L$ increases with $L$ until its maximal possible value $R_{L_\text{c}}\equiv R_\infty$. Here, the training data is overcomplete and sufficient to learn all degrees of freedom of $U(\boldsymbol{\theta})$:
\begin{defn}[Overcomplete data]\label{eq:ovp_S}
A given model $U(\boldsymbol{\theta})$ and $\rho_L$ is overcomplete  when $R_L$ does not increase further upon increasing $L$. Its maximal rank $R_\infty= R_{L_\text{c}}$ is reached for a critical number of training data $L\ge L_\text{c}$
\begin{equation}
R_\infty = R_{L_\text{c}}\equiv\max_{L\ge L_\text{c}}R_L(\rho_L,U)\,.
\end{equation}
\end{defn}

We bound $R_L$ similar to $R_1$ for Ref.~\cite{larocca2021theory} (see SM~\ref{sec:liebound}):
\begin{theorem}\label{th:liebound}
    The maximal rank $R_L$  is bounded by the dimension of the DLA
        $R_L\leq\dim(\mathfrak{g})$ 
    where $\mathfrak{g}=\mathrm{span}\left\langle iH_1, \ldots, iH_K \right\rangle_\text{Lie}$ is generated by the repeated nested commutators of the generators $H_k$ of $U(\boldsymbol{\theta})$.
\end{theorem}
Thus, using an ansatz with restricted Lie algebra~\cite{wiersema2023classification,nguyen2022theory,meyer2022exploiting} with $\text{dim}(\mathfrak{g})\sim\text{poly}(N)$ generalizes with $L_\text{c},M_\text{c}\sim\text{poly}(N)$ where $N$ is the number of qubits.

We can estimate $L_\text{c}$ with the following consideration: To generalize we have to learn all $R_\infty$ degrees of freedom of the unitary. The first training state allows us to learn $R_1$ degrees of freedom, while each additional state provides a bit less as seen in~\eqref{eq:DLsin}. For the upper bound of~\eqref{eq:DLsin} we have $L_\text{c}\approx 2 R_\infty/R_1$, which we numerically find to be a good estimator also for other models:
\begin{observation}[Generalization for learning unitaries]\label{obs:threshold}
A trained model generalizes $C_\text{test}(\boldsymbol{\theta}^*)\approx0$ with high probability when the model is overparameterized (i.e. $M\ge M_\text{c}\ge R_L$ for Def.~\ref{def:ovp_U}) and overcomplete (i.e. $L\ge L_\text{c}$ for Def.~\ref{eq:ovp_S}). The critical number of training states $L_\text{c}$ needed to generalize can be approximated by
\begin{equation}\label{eq:Lcapprox}
    L_\text{c}\approx 2R_\infty/R_1\,.
\end{equation}
\end{observation}

\prlsection{Applications}
We want to learn the unitary evolution $V_{XY}=\exp(-iH_{XY} t)$ at time $t$ of the XY-Hamiltonian $H_{XY}=\sum_{k=1}^N(\sigma_k^x\sigma_{k+1}^x+\sigma_k^y\sigma_{k+1}^y+h_k \sigma_k^z)$, where $\sigma^\alpha_k$, $\alpha\in\{x,y,z\}$ is the Pauli operator acting on qubit $k$  and  $h_k\in \mathbb{R}$. We learn $V_{XY}$ with $U_{XY}(\boldsymbol{\theta})$ ansatz (see SM~\ref{sec:ansatz} for definition), which can represent any $V_{XY}$ with polynomial number of parameters~\cite{kokcu2022algebraic,kokcu2022fixed}. $H_{XY}$ and $U_{XY}$ conserve the particle number operator $P=\sum_{k=1}^N\frac{1}{2}(1-\sigma^z_k)$ with $[U_{XY},P]=[H_{XY},P]=0$, where $[.]$ is the commutator. As training states, we use $\ket{\psi_{\ell}}\in W_{p=1}$ which are states symmetric in regards to $P$, i.e. $P\ket{\psi_{\ell}}=p\ket{\psi_{\ell}}$  with the same eigenvalue $p=1$ for all $\ket{\psi_{\ell}}\in W_{p=1}$. Further, we have the single-qubit product states $W_{\text{prod}}$ with $\ket{\psi_\ell}=\bigotimes_{k=1}^N\ket{\phi^{k}_\ell},\, \ket{\phi^{k}_\ell}\in\mathcal{H}(\mathbb{C}^2)$ which are not symmetric in respect to $P$.
\begin{observation}[Non-symmetric data improves generalization]\label{obs:symmetry}
We train $U_{\text{XY}}(\boldsymbol{\theta})$ with $(i)$ particle-number conserving states  $\ket{\psi_\ell}\in W_{p=1}$ 
and $(ii)$ single-qubit product states $\ket{\psi_\ell}\in W_{\text{prod}}$. 
\revA{For $W_{p=1}$ we find exactly $R_1=2N-2$, $R_\infty=N^2-1$, while for $W_{\mathrm{prod}}$ we find via numerical extrapolation $R_1=2N^2-3N+2$ and $R_\infty=2N^2-1$ (Fig.~\ref{fig:rank}c,d). Generalization requires less $L_\text{c}\approx 2R_\infty/R_1$ training states for non-symmetric data:}
\begin{align*}
(i) \text{\phantom{Non-}Symmetric}: \; L_\text{c}=&N \;\mathrm{for}\; \ket{\psi_\ell}\in W_{p=1}\\
(ii) \text{Non-symmetric}:\;L_\text{c}=&2 \;\;\mathrm{for}\;  \ket{\psi_\ell}\in W_{\mathrm{prod}}\,\,,N>4\,.
\end{align*}
\end{observation}
\revA{Intuitively, non-symmetric data requires less $L$ as it can use information from other symmetry sectors.}

Next, we consider out-of-distribution generalization where the training data is drawn from a different distribution than the test data~\cite{caro2022out}: 
\begin{observation}[Out-of-distribution generalization requires less data]\label{obs:outofdist}
Training $U_{XY}(\boldsymbol{\theta})$ with product states $\ket{\psi_\ell}\in W_{\mathrm{prod}}$, but testing with number-conserving data $W_{p=1}$ achieves out-of-distribution generalization with only $L\ge 2$ training data.
In contrast, in-distribution training and testing with number-conserving data $\ket{\psi_\ell}\in W_{p=1}$ requires $L\ge N$ states to generalize.
\end{observation}
\revA{This result follows from Obs.~\ref{obs:symmetry} and product states being sufficient to learn arbitrary unitaries~\cite{caro2022out}. We confirm our result numerically in Fig.~\ref{fig:test}d}

\prlsection{Numerical results}
In Fig.~\ref{fig:test}a-c we study learning of unitaries with hardware-efficient ansatz $U_\text{HE}(\boldsymbol{\theta})$ (see SM~\ref{sec:ansatz}). In Fig.~\ref{fig:test}a, we converge to local minima with $C_\text{train}\gg0$ for $M\le R_L$, while we find global minimum $C_\text{train}\approx0$ for $M\ge R_L$, which is indicated as black dashed line. In Fig.~\ref{fig:test}b, generalization $C_\text{test}\approx0$ is achieved only for $M\ge R_\infty$ and $L\ge L_\text{c}\approx 2R_\infty/R_1$ indicated by the vertical black line.
In Fig.~\ref{fig:test}c, the number of training steps $E$ to converge show characteristic peaks close to $M_\text{c}$ and $L_\text{c}$ indicated by black dashed lines. 
\revA{In Fig.~\ref{fig:test}d we show the test error against $L$ for the $U_{XY}$ ansatz which conserves particle number $P$. We find that training with symmetric data $W_{\mathrm{prod}}$ generalizes for $L\ge2$, while training with non-symmetric $W_{p=1}$ generalizes for $L\ge N$ which numerically confirms Obs.~\ref{obs:symmetry}. Further, the green curve shows out-of-distribution generalization where training with $W_{\mathrm{prod}}$ generalizes with test data from  $W_{p=1}$ using only $L\ge2$, while in-distribution learning (blue curve) requires $L\ge N$, confirming Obs.~\ref{obs:outofdist}.
We study $U_\text{XY}$ in more detail in SM~\ref{sec:trainXY} and other models which generalize for constant $L$ in SM~\ref{sec:training_gen}.}

\prlsection{Conclusion}
Our newly introduced DQFIM $\mathcal{Q}$ and its maximal rank $R_L$ quantify the learnable degrees of freedom of ansatz $U(\boldsymbol{\theta})$ using $L$ training states.  $R_L$ increases with $L$ until the training data becomes overcomplete at $R_{L_\text{c}}=R_\infty$ and $L_\text{c}\approx 2R_\infty/R_1$ where one is able generalize. 

Overparameterized models converge to global minima with high probability~\cite{rabitz2004quantum,larocca2021theory,you2022convergence,anschuetz2021critical,kim2021universal,campos2021abrupt,kim2022quantum}. We show that overparameterization depends on $L$ and occurs for $M\ge M_\text{c}(L)\ge R_L$ circuit parameters. 
Overparameterization and generalization appear in three distinct regimes, where training time increases substantially at the transitions, potentially indicating computational phase transitions~\cite{kiani2020learning,anschuetz2021critical}.

While symmetries have been shown to improve generalization~\cite{meyer2022exploiting,larocca2022group}, we show that symmetries in data can also \emph{increase $L$ needed to generalize} due to higher $R_\infty/R_1$ ratio compared to non-symmetric data. 
This also implies that \emph{out-of-distribution generalization can outperform in-distribution generalization} when training on non-symmetric data, but testing on symmetric data.
Note that non-symmetric data has larger $M_\text{c}$, which implies an interesting trade-off between $L_\text{c}$ and $M_\text{c}$.

\revA{The DQFIM accurately characterizes overparameterization and generalization depending on the specific structure and symmetries of ansatz $U(\boldsymbol{\theta})$ and training data $\rho_L$. 
In contrast, previously considered uniform generalization bounds 
provide only a loose bound on generalization error $\sim \sqrt{1/L}$ without accounting for symmetries~\cite{caro2021generalization,banchi2021generalization}. 
We demonstrate the relationship between DLA and generalization, showing that polynomial DLA implies overparameterization and generalization with polynomial circuit depth and dataset size. Generalization with few data is possible whenever $R_\infty/R_1=\text{const}$, explaining the numerical observations of Ref.~\cite{caro2021generalization,gibbs2022dynamical} (see also SM~\ref{sec:training_gen}). Thus, problem classes with polynomial DLA~\cite{wiersema2023classification,schatzki2022theoretical,sauvage2024building} can be trained with low data cost and avoid barren plateaus~\cite{fontana2023adjoint,ragone2023unified}.  }

\revA{Our results apply to various quantum machine learning algorithms. We study unitary learning problems, which includes quantum compiling~\cite{khatri2019quantum}, quantum control (SM~\ref{sec:control}), and quantum generative models (SM~\ref{sec:generative}). In SM~\ref{sec:excited} the DQFIM determines convergence of the subspace-search variational quantum eigensolver (SSVQE) for finding eigenstates of Hamiltonians~\cite{nakanishi2019subspace}. In SM~\ref{sec:classify} we apply the DQFIM for classification tasks. Here, the label operator $O_\ell=y_\ell\sigma^z$ is not a projector and thus has not only one, but $2^{N-1}$ degenerate solutions. This reduces $M_\text{c}(L)\approx R_L \gamma$ by a constant factor $\gamma\leq1$ where for a generic ansatz we find $\gamma=\frac{1}{2}$. $\gamma$ can be smaller when the ansatz has symmetries which opens an interesting approach to reduce circuit depth in classification tasks.}

Numerical evaluation of the DQFIM is straightforward via differentiation (with code available online~\cite{haug2023gen}) and is scalable for matrix product states. Quantum computers can efficiently measure the DQFIM using the Hadamard test with a single ancilla and two control operations, or alternatively the shift-rule and purification~\cite{mari2021estimating} in SM~\ref{sec:dQFIM}.

While the complexity of unitaries grows linearly with $M$~\cite{haferkamp2022linear,haug2021capacity}, we find that the learnable degrees of freedom $R_L$ of unitaries grows only sub-linearly with $L$. 
Generalization error for overparameterized models scales as $C_\text{test}\sim 1- (L/L_\text{c})^2$ (see SM~\ref{sec:trainXY}) which saturates the lower bound derived in Ref.~\cite{sharma2022reformulation}. We note that for underparameterized models the empirical  generalization error $C_\text{test}-C_\text{train}$~\cite{gil2023understanding} is not a good indicator of generalization due to convergence to bad local minima (see SM~\ref{sec:risk}).

Finally, future work can apply the DQFIM for kernel models~\cite{schuld2019quantum,haug2021large}, noisy systems~\cite{garcia2023effects} and \revB{quantum natural gradients~\cite{stokes2020quantum}.}

 \let\oldaddcontentsline\addcontentsline
\renewcommand{\addcontentsline}[3]{}

\medskip
\begin{acknowledgments}
{{\em Acknowledgements---}} We thank Adithya Sireesh and Raghavendra Peddinti for discussions.
This work is supported by a Samsung GRC project and the UK Hub in Quantum Computing and Simulation, part of the UK National Quantum Technologies Programme with funding from UKRI EPSRC grant EP/T001062/1. 
\end{acknowledgments}
\bibliography{generalization}

\let\addcontentsline\oldaddcontentsline

\onecolumngrid
\newpage 

\appendix
\setcounter{secnumdepth}{2}


\clearpage
\begin{center}
	\textbf{\large Supplemental Material}
\end{center}
\makeatletter
In the Supplemental Material, we derive the data quantum Fisher information metric (DQFIM) from first principles, connect it to the purification of the quantum Fisher information metric, and show how to measure the DQFIM on quantum computers. 
We prove that the dynamical Lie algebra  bounds the DQFIM. Further, we define the ansatz unitaries used in the main text and study the DQFIM with other types of variational ansatze. Finally, we use the DQFIM to study overparameterization and generalization for various quantum machine learning models, such as quantum control, generative models, finding eigenstates of Hamiltonians and classification tasks.

\makeatletter
\@starttoc{toc}

\makeatother

\section{Data quantum Fisher information (DQFIM)}\label{sec:dQFIM}
Here, we derive our data quantum Fisher information (DQFIM) from first principles.
First, we review the quantum Fisher information (QFIM). Then, we derive the DQFIM using two different approaches: First, as metric of unitary evolution mapped onto the subspace of the training data, and second as the QFIM of the purification of the training data. 
Finally, we show how to compute the DQFIM using two methods: Either by using the Hadamard test using one ancilla and two controlled operations, or by using the shift-rule on the purification.

\subsection{Recap of quantum Fisher information metric}
The Quantum Fisher information metric (QFIM) ~\cite{meyer2021fisher,liu2020quantum} is an essential tool for quantum sensing, parameter estimation and optimization of quantum circuits.
Here, we review the derivation of the QFIM or Fubini-Study metric $\mathcal{F}$~\cite{cheng2010quantum}.
We have a parameterized quantum state $\ket{\psi(\boldsymbol{\theta})}$.
 We now study the variation
\begin{align*}
&\text{d}s^2=\vert\vert \ket{\psi(\boldsymbol{\theta}+\text{d}\boldsymbol{\theta})}-\ket{\psi(\boldsymbol{\theta})}\vert\vert= \braket{\delta\psi}{\delta\psi}=\\
&\braket{\partial_n \psi}{ \partial_m \psi}\text{d}\boldsymbol{\theta}^n\boldsymbol{\theta}^m=(\gamma_{nm}+i\sigma_{nm})\text{d}\boldsymbol{\theta}^n\text{d}\boldsymbol{\theta}^m
\end{align*}
where we defined the real and imaginary part  $\gamma_{nm}$ and $\sigma_{nm}$ of $\braket{\partial_n \psi}{ \partial_m \psi}$. As $\text{d}s^2$ is hermitian, we have $\gamma_{nm}=\gamma_{mn}$ and $\sigma_{nm}=-\sigma_{mn}$, such that $\sigma_{nm}\text{d}\boldsymbol{\theta}^n\text{d}\boldsymbol{\theta}^m$ vanishes. However, $\gamma_{nm}$ is not a proper metric as it is not invariant under the gauge transformation $\ket{\psi'}=\exp(i\alpha)\ket{\psi}$ under a global phase rotation with $\alpha$. We now construct a proper gauge invariant metric.

First, one can easily show by using $\braket{\psi}{\psi}=1$ that $\beta_n=i\braket{\psi}{ \partial_n \psi}\in\mathbb{R}$. 
Next, we compute $\braket{\psi'}{\psi'}=\gamma_{nm}'+i\sigma_{nm}'$, where a straightforward calculation yields
\begin{equation}
\gamma_{nm}'=\gamma_{nm}+\partial_n\alpha\partial_m\alpha-\beta_m\gamma_n\alpha-\beta_n\gamma_m
\end{equation}
and $\sigma_{nm}'=\sigma_{nm}$.
From this result, we now define a gauge-invariant metric
\begin{equation}
g_{nm}=\gamma_{nm}-\beta_n\beta_m
\end{equation}
where one can easily confirm $g_{nm}'=g_{nm}$ by using $\beta_n'=\beta_n-\partial_n\alpha$.
We can think of $\gamma_{nm}$ measuring the change of $\ket{\psi(\boldsymbol{\theta})}$ in the Hilbertspace, while $g_{nm}$ measures its change excluding global phases which have no observable effect.

The quantum geometric tensor is defined as
\begin{equation}
\Xi_{nm}=\braket{\partial_n \psi}{ \partial_m \psi}-\braket{\partial_n\psi}{\psi}\braket{\psi}{\partial_m\psi}
\end{equation}
and the QFIM as $\mathcal{F}_{nm}=4\text{Re}(\Xi_{nm})$, which corresponds to the real part of $g_{nm}$.

The QFIM describes the change in fidelity for $\vert\braket{\psi(\boldsymbol{\theta})}{\psi(\boldsymbol{\theta}+\text{d}\boldsymbol{\theta})}\vert^2$ as we will see in the following.
First, note that $\braket{\psi}{\partial_n\psi}\in\text{Im}$. Thus, its derivative must be also imaginary, i.e. $\braket{\psi}{\partial_n\partial_m\psi}+\braket{\partial_m\psi}{\partial_n\psi}\in\text{Im}$, which immediately implies 
\begin{equation}\label{eq:helpU_fub}
\braket{\psi}{\partial_n\partial_m\psi}=-\braket{\partial_m\psi}{\partial_n\psi}\,.
\end{equation}
Now, we find via Taylor expansion
\begin{align*}
&\braket{\psi(\boldsymbol{\theta})}{\psi(\boldsymbol{\theta}+\text{d}\boldsymbol{\theta})}\approx\numberthis\label{eq:Taylor_fub}\\
&1+i\braket{\psi}{\partial_n\psi}\text{d}\boldsymbol{\theta}^n
+\frac{1}{2}\braket{\psi}{\partial_n\partial_m\psi}\text{d}\boldsymbol{\theta}^n\text{d}\boldsymbol{\theta}^m=\\
&1+i\braket{\psi}{\partial_n\psi}\text{d}\boldsymbol{\theta}^n
-\frac{1}{2}\braket{\partial_n\psi}{\partial_m\psi}\text{d}\boldsymbol{\theta}^n\text{d}\boldsymbol{\theta}^m
\end{align*}
Now, we compute using~\eqref{eq:Taylor_fub}
\begin{align*}
&\vert\braket{\psi(\boldsymbol{\theta})}{\psi(\boldsymbol{\theta}+\text{d}\boldsymbol{\theta})}\vert^2=\\
&\braket{\psi(\boldsymbol{\theta})}{\psi(\boldsymbol{\theta}+\text{d}\boldsymbol{\theta})}\braket{\psi(\boldsymbol{\theta}+\text{d}\boldsymbol{\theta})}{\psi(\boldsymbol{\theta})}\approx\\
&1-[\braket{\partial_n\psi}{\psi}\braket{\psi}{\partial_m\psi}\\
&+\frac{1}{2}\braket{\partial_n\psi}{\partial_m\psi}+\frac{1}{2}\braket{\partial_m\psi}{\partial_n\psi}]\text{d}\boldsymbol{\theta}^n\text{d}\boldsymbol{\theta}^m=\\
&1-\text{Re}[\braket{\partial_n\psi}{\partial_m\psi}-\braket{\partial_n\psi}{\psi}\braket{\psi}{\partial_m\psi}]\text{d}\boldsymbol{\theta}^n\text{d}\boldsymbol{\theta}^m.
\end{align*}
where in the second step we used \eqref{eq:helpU_fub}.
Finally, we have
\begin{equation}
\vert\braket{\psi(\boldsymbol{\theta})}{\psi(\boldsymbol{\theta}+\text{d}\boldsymbol{\theta})}\vert^2=1-\frac{1}{4}\mathcal{F}_{nm}\text{d}\boldsymbol{\theta}^n\text{d}\boldsymbol{\theta}^m\,.
\end{equation}
This implies that $\mathcal{F}$ is a metric that describes the change in state space under a variation in parameter $\boldsymbol{\theta}$.

\subsection{Derivation of DQFIM}

We now generalize the QFIM to the DQFIM, which describes learning with $L$ training states. We have a training set $S_L=\{\ket{\psi_{\ell}},O_\ell\}_{\ell=1}^L$ and an ansatz $U(\theta)\equiv U$.
We recall the datastate 
\begin{equation}
\rho_L=L^{-1}\sum_{\ell=1}^L\ket{\psi_{\ell}}\bra{\psi_{\ell}}
\end{equation}
which is the mixture of all training states.

We now derive the DQFIM~\eqref{eq:Fisher_sup} from a variational principle in a similar manner as the QFIM.
We consider $U(\boldsymbol{\theta})$ mapped onto the subspace of the datastate $\rho_L$ written as $U(\boldsymbol{\theta})\rho_L$. The degrees of freedom of $U(\boldsymbol{\theta})\rho_L$ describe the degrees of freedom that can be learned about $U(\boldsymbol{\theta})$ using the training dataset. 

We now have
\begin{align*}
&\text{d}s^2=\vert\vert (U(\boldsymbol{\theta}+\text{d}\boldsymbol{\theta})-U(\boldsymbol{\theta}))\sqrt{\rho_L}\vert\vert/\beta=\text{tr}(\delta U\rho_L\delta U^\dagger)=\\
&\text{tr}(\partial_n U \rho_L\partial_m U^\dagger)\text{d}\boldsymbol{\theta}^n\boldsymbol{\theta}^m=(\gamma_{nm}+i\sigma_{nm})\text{d}\boldsymbol{\theta}^n\text{d}\boldsymbol{\theta}^m\,,
\end{align*}
where we have the difference $\delta U=U(\boldsymbol{\theta}+\text{d}\boldsymbol{\theta})-U(\boldsymbol{\theta})$, the square of the Frobenius norm $\vert\vert A\vert\vert=\text{tr}(A^\dagger A)$, the real and imaginary part  $\gamma_{nm}$ and $\sigma_{nm}$ of $\text{tr}(\partial_n U\rho_L \partial_m U^\dagger)$.
Note that we have $\text{tr}(U \rho_L U^\dagger)=1$. One can now immediately check that one recovers the regular QFIM for $\rho_1=\ket{\psi}\bra{\psi}$.
As $\text{d}s^2$ is hermitian, we have $\gamma_{nm}=\gamma_{mn}$ and $\sigma_{nm}=-\sigma_{mn}$, such that $\sigma_{nm}\text{d}\boldsymbol{\theta}^n\text{d}\boldsymbol{\theta}^m$ vanishes. However, $\gamma_{nm}$ is not a proper metric as it is not invariant under the gauge transformation $U'=\exp(i\alpha)U$, i.e. a global phase rotation with $\alpha$. We now construct a proper gauge invariant metric.

First, we apply $\partial_n$ to $\text{tr}(U\rho_L U^\dagger)=1$ and see that $\text{tr}(\partial_n U\rho_L U^\dagger)+\text{tr}(U\rho_L \partial_n U^\dagger)=0$. It follows that $\text{tr}(U\rho_L \partial_n U^\dagger)+\text{tr}(U\rho_L \partial_n U^\dagger)^\dagger=0$ and thus  $\beta_n=i\text{tr}(U\rho_L\partial_n U^\dagger)\in\mathbb{R}$. 
Next, we compute $\text{tr}({U'} \rho_L {U'}^\dagger)=\gamma_{nm}'+i\sigma_{nm}'$, where a straightforward calculation yields
\begin{equation}
\gamma_{nm}'=\gamma_{nm}+\partial_n\alpha\partial_m\alpha-\beta_m\gamma_n\alpha-\beta_n\gamma_m
\end{equation}
and $\sigma_{nm}'=\sigma_{nm}$.
From this result, we now define a gauge-invariant metric
\begin{equation}
g_{nm}=\gamma_{nm}-\beta_n\beta_m
\end{equation}
where one can easily confirm $g_{nm}'=g_{nm}$ by using $\beta_n'=\beta_n-\partial_n\alpha$.
We can think of $\gamma_{nm}$ measuring the change of $U(\boldsymbol{\theta})\rho_L$ in the full isometric space, while $g_{nm}$ measures the change excluding global phases which have no observable effect.

In analogy to the quantum geometric tensor, we define the data quantum geometric tensor 
\begin{equation}
\Xi_{nm}=\text{tr}(\partial_n U\rho_L \partial_m U^\dagger)-\text{tr}(\partial_n U\rho_L U^\dagger)\text{tr}(U \rho_L\partial_m U^\dagger)
\end{equation}
and the DQFIM as $\mathcal{Q}_{nm}=4\text{Re}(\Xi_{nm})$, which corresponds to the real part of $g_{nm}$, with
\begin{align*}
&\mathcal{Q}_{nm}(\rho_L)=
4\text{Re}(\text{tr}(\partial_n U \rho_L \partial_m U^\dagger)-\text{tr}(\partial_n U\rho_L U^\dagger)\text{tr}(U \rho_L\partial_m U^\dagger))\,.\numberthis\label{eq:Fisher_sup}
\end{align*}
Indeed, we find for $L=1$ that $\mathcal{Q}(\rho_1)=\mathcal{F}$. In contrast, for a training set with $\rho_L=I/d$, we find what we call the unitary QFIM
\begin{align*}
&\mathcal{Q}_{nm}^I\equiv\mathcal{Q}_{nm}(I/d)=
4\text{Re}(d^{-1}\text{tr}(\partial_n U \partial_m U^\dagger)-d^{-2}\text{tr}(\partial_n U U^\dagger)\text{tr}(U \partial_m U^\dagger))\,.\numberthis\label{eq:Fisher_unitary}
\end{align*}
The DQFIM describes the change of $\vert\text{tr}(U(\boldsymbol{\theta}) \rho_L U^\dagger(\boldsymbol{\theta}+\text{d}\boldsymbol{\theta}))\vert^2$ which we are going to show in the following.
First, note that $\text{tr}(U\rho_L\partial_n U^\dagger)\in\text{Im}$. Thus, its derivative must be also imaginary, i.e. $\text{tr}(U\rho_L\partial_n\partial_m U^\dagger)+\text{tr}(\partial_nU\rho_L\partial_m U^\dagger)\in\text{Im}$, which immediately implies 
\begin{equation}\label{eq:helpU}
\text{tr}(U\rho_L\partial_n\partial_m U^\dagger)=-\text{tr}(\partial_nU\rho_L\partial_m U^\dagger)\,.
\end{equation}
Now, we find via Taylor expansion
\begin{align*}
&\text{tr}(U(\boldsymbol{\theta}) \rho_L U(\boldsymbol{\theta}+\text{d}\boldsymbol{\theta})^\dagger)\approx\numberthis\label{eq:Taylor}\\
&1+i\text{tr}( U \rho_L\partial_n U^\dagger)\text{d}\boldsymbol{\theta}^n
+\frac{1}{2}\text{tr}(U\rho_L \partial_n\partial_m U^\dagger)\text{d}\boldsymbol{\theta}^n\text{d}\boldsymbol{\theta}^m=\\
&1+i\text{tr}( U\rho_L \partial_n U^\dagger)\text{d}\boldsymbol{\theta}^n
-\frac{1}{2}\text{tr}(\partial_nU\rho_L \partial_m U^\dagger)\text{d}\boldsymbol{\theta}^n\text{d}\boldsymbol{\theta}^m
\end{align*}
Now, we compute using~\eqref{eq:Taylor}
\begin{align*}
&\vert\text{tr}(U(\boldsymbol{\theta}) \rho_L U(\boldsymbol{\theta}+\text{d}\boldsymbol{\theta})^\dagger)\vert^2=\\
&\text{tr}(U(\boldsymbol{\theta}) \rho_L U(\boldsymbol{\theta}+\text{d}\boldsymbol{\theta})^\dagger)\text{tr}(U(\boldsymbol{\theta}+\text{d}\boldsymbol{\theta}) U(\boldsymbol{\theta})^\dagger)\approx\\
&1-[\text{tr}(\partial_n U\rho_L U^\dagger)\text{tr}( U\rho_L \partial_mU^\dagger)\\
&+\frac{1}{2}\text{tr}(\partial_nU \rho_L\partial_m U^\dagger)+\frac{1}{2}\text{tr}(\partial_m U\rho_L  \partial_n U^\dagger)]\text{d}\boldsymbol{\theta}^n\text{d}\boldsymbol{\theta}^m=\\
&1-\text{Re}[\text{tr}(\partial_nU\rho_L \partial_m U^\dagger)-\\
&\text{tr}(\partial_n U\rho_L U^\dagger)\text{tr}(U \partial_m\rho_L U^\dagger)]\text{d}\boldsymbol{\theta}^n\text{d}\boldsymbol{\theta}^m.
\end{align*}
where in the second step we used \eqref{eq:helpU}.
Finally, we have
\begin{equation}
\vert\text{tr}(U(\boldsymbol{\theta}) \rho_L U(\boldsymbol{\theta}+\text{d}\boldsymbol{\theta})^\dagger)\vert^2=1-\frac{1}{4}\mathcal{Q}_{nm}\text{d}\boldsymbol{\theta}^n\text{d}\boldsymbol{\theta}^m\,,
\end{equation}
which relates a change in parameter $\boldsymbol{\theta}$ to the change in the unitary mapped onto the training states.

\revA{
\subsection{DQFIM as purification of QFIM}\label{sec:dQFIM_meas}
We now show an alternative derivation of the DQFIM as the QFIM of the purification of the dataset.

The DQFIM is equivalent to the QFIM of the purification of the data state $\rho_L=\frac{1}{L}\sum_{i=1}^L \ket{\psi_i}\bra{\psi_i}$. In particular,  the purification $\ket{\chi_L}$ of $\rho_L$ can be written as
\begin{equation}
    \ket{\chi_L}=\frac{1}{\sqrt{L}} \sum_{i=1}^L\ket{\psi_i}\ket{i}
\end{equation}
where the data state is the partial trace $\text{tr}_B(.)$ over the $L$ ancilla degrees of freedom $\ket{i}$
\begin{equation}
    \rho_L= \text{tr}_B(\ket{\chi_L}\bra{\chi_L})=\sum_{i=1}^L (I_N\otimes \bra{i}) \ket{\chi_L}\bra{\chi_L} (I_N\otimes \ket{i})=\frac{1}{L} \sum_{i=1}^L\ket{\psi_i}\bra{\psi_i}\,.
\end{equation}
We now define the evolved purified state $U(\boldsymbol{\theta})\otimes I_B\ket{\chi_L}$ where $I_B$ with $\text{dim}(I_B)=\log_2(L)$ is the identity over the ancilla qubits of the purfication degrees of freedom. A straightforward calculation shows that the QFIM $\mathcal{F}_{nm}$ of the purified data state is equivalent to the DQFIM $\mathcal{Q}_{nm}(\rho_L,U(\boldsymbol{\theta}))$ via
\begin{align*}
    &\mathcal{F}_{nm}(U(\boldsymbol{\theta})\otimes I_B\ket{\chi_L})=\\
    &4\text{Re}(\bra{\chi_L}\partial_n U(\boldsymbol{\theta})^\dagger\otimes I_B \partial_m U(\boldsymbol{\theta})\otimes I_B\ket{\chi_L}-\bra{\chi_L} U(\boldsymbol{\theta})^\dagger\otimes I_B \partial_n U(\boldsymbol{\theta})\otimes I_B\ket{\chi_L} \bra{\chi_L} \partial_m U(\boldsymbol{\theta})^\dagger\otimes I_B  U(\boldsymbol{\theta})\otimes I_B\ket{\chi_L}=\\
    &4\text{Re}(\frac{1}{L}\sum_{ij} \bra{\psi_i}\bra{i} \partial_n U(\boldsymbol{\theta})^\dagger\otimes I_B \partial_m U(\boldsymbol{\theta})\otimes I_B \ket{\psi_j}\ket{j}-\\
    &\frac{1}{L^2}\sum_{ij} \bra{\psi_i}\bra{i} U(\boldsymbol{\theta})^\dagger\otimes I_B \partial_n U(\boldsymbol{\theta})\otimes I_B\ket{\psi_j}\ket{j}
    \sum_{ij}\bra{\psi_i}\bra{i} \partial_m U(\boldsymbol{\theta})^\dagger\otimes I_B  U(\boldsymbol{\theta})\otimes I_B\ket{\psi_j}\ket{j})= \\
    & 4\text{Re}(\frac{1}{L}\sum_{i} \bra{\psi_i} \partial_n U(\boldsymbol{\theta})^\dagger \partial_m U(\boldsymbol{\theta}) \ket{\psi_i}-\frac{1}{L^2}\sum_{i} \bra{\psi_i} U(\boldsymbol{\theta})^\dagger \partial_n U(\boldsymbol{\theta})\ket{\psi_i}
    \sum_i\bra{\psi_i} \partial_m U(\boldsymbol{\theta})^\dagger  U(\boldsymbol{\theta})\ket{\psi_i})=\\
    & 4\text{Re}(\text{tr}(\frac{1}{L}\sum_{i} \ket{\psi_i}\bra{\psi_i} \partial_n U(\boldsymbol{\theta})^\dagger \partial_m U(\boldsymbol{\theta})) -\text{tr}(\frac{1}{L}\sum_{i} \ket{\psi_i}\bra{\psi_i} U(\boldsymbol{\theta})^\dagger \partial_n U(\boldsymbol{\theta}))
    \text{tr}(\sum_i \frac{1}{L}\ket{\psi_i}\bra{\psi_i} \partial_m U(\boldsymbol{\theta})^\dagger  U(\boldsymbol{\theta})))=
    \\
    & 4\text{Re}(\text{tr}( \partial_m U(\boldsymbol{\theta}) \rho_L \partial_n U(\boldsymbol{\theta})^\dagger ) -\text{tr}( \partial_n U(\boldsymbol{\theta}) \rho_L U(\boldsymbol{\theta})^\dagger )
    \text{tr}( U(\boldsymbol{\theta})\rho_L \partial_m U(\boldsymbol{\theta})^\dagger  ))=\mathcal{Q}_{nm}(\rho_L, U(\boldsymbol{\theta}))\,.
\end{align*}

We now  propose two measurement scheme for the DQFIM for experiments on quantum computers.
\subsection{Measuring DQFIM on quantum computers using Hadamard test}\label{sec:dQFIM_meas_h}
We now propose an approach to measure the DQFIM using the Hadamard test.

First note that that our ansatz~\eqref{eq:ansatz_sup} consist of $G$ layers with 
\begin{equation}
    U(\boldsymbol{\theta})=\prod_{k=1}^G U_k(\boldsymbol{\theta}_k)\,\,\text{with}\,\, U^{(k)}(\theta_k)=\prod_{n=1}^K\exp(-i\theta_{kn} H_n)
\end{equation}
where $U^{(k)}(\theta_k)$ is the unitary of the $k$th layer. Here, $H_n$ are $K$ hermitian matrices and $\boldsymbol{\theta}_k=\{\theta_{k1},\dots,\theta_{kn}\}$ are the parameters of the $k$th layer. The total parameter vector $\theta=\{\boldsymbol{\theta}_1,\dots,\boldsymbol{\theta}_G\}$ of the ansatz has $M=GK$ parameters.

As we have $\partial_n \exp(-i\theta_n H_n)= -iH_n\exp(-i\theta_n H_n)$, we can write the derivatives as 
\begin{equation}
\partial_n U(\boldsymbol{\theta})=-iU_{n+1\rightarrow M}H_nU_{1\rightarrow n}\,.
\end{equation}
Here, we define  
\begin{equation}
U_{m\rightarrow n}=U_n U_{n-1} \dots U_{m+1}U_m
\end{equation}
as the propagator from layer $m$ to layer $n$.

We have the DQFIM with
\begin{align*}
&\mathcal{Q}_{nm}(\rho_L)=
4\text{Re}(\text{tr}(\partial_n U \rho_L \partial_m U^\dagger)-\text{tr}(\partial_n U\rho_L U^\dagger)\text{tr}(U \rho_L\partial_m U^\dagger))\,.
\end{align*}

First, computing $\text{tr}(U \rho_L\partial_m U^\dagger))$ and $\text{tr}(\partial_n U\rho_L U^\dagger)$ is straightforward as they are observables. In particular, using $\rho_L=\frac{1}{L}\sum_{i=1}^L \ket{\psi_i}\bra{\psi_i}$ and~\eqref{eq:deriv} we have
\begin{align*}
    &\text{tr}(\partial_n U \rho_L U^\dagger))=\frac{1}{L}\sum_{i=1}^L \bra{\psi_i}U^\dagger \partial_n U\ket{\psi_i}=\frac{1}{L}\sum_{i=1}^L \bra{\psi_i}U^\dagger U_{n+1\rightarrow M} (-iH_n)U_{1\rightarrow n}\ket{\psi_i}=\frac{-i}{L}\sum_{i=1}^L \bra{\psi_i}U_{1\rightarrow n}^\dagger H_n U_{1\rightarrow n}\ket{\psi_i}
\end{align*}
i.e. one needs to simply evolve each training state with $U_{1\rightarrow n}\ket{\psi_i}$ and measure the operator $H_n$. Similarly, one measures $\text{tr}(U\rho_L \partial_m U^\dagger)$.

Next, we measure $\text{Re}(\text{tr}(\partial_n U \rho_L \partial_m U^\dagger))$ which is not an observable, but can be efficiently measured using the Hadamard test~\cite{li2017efficient,yuan2019theory}. In particular, assuming $n>m$ we have
\begin{align*}
&\text{Re}(\text{tr}(\partial_n U \rho_L \partial_m U^\dagger))=\frac{1}{L}\sum_{i=1}^L \text{Re}(\bra{\psi_i}\partial_m U^\dagger \partial_n U\ket{\psi_i})=\\
&\frac{1}{L}\sum_{i=1}^L \text{Re}(\bra{\psi_i}U_{1\rightarrow n}^\dagger (iH_m) U_{m+1\rightarrow M}^\dagger U_{n+1\rightarrow M} (-iH_n)U_{1\rightarrow n}\ket{\psi_i})=\frac{1}{L}\sum_{i=1}^L \text{Re}(\bra{\psi_i}U_{1\rightarrow m}^\dagger H_m U_{m+1\rightarrow n}^\dagger H_n U_{1\rightarrow n}\ket{\psi_i})
\end{align*}
Terms of the form $\text{Re}(\bra{\psi_i}U_{1\rightarrow m}^\dagger H_m U_{m+1\rightarrow n}^\dagger H_n U_{1\rightarrow n}\ket{\psi_i})$ can be efficiently measured using the Hadamard test described in~\cite{yuan2019theory}. It requires one ancilla qubit, and replacing $H_m$, $H_n$ with controlled operators, e.g. $H_n=\sigma^x_1$ becomes a CNOT with control on the ancilla and target on the first qubit. 
So in total,  we require a single ancilla and two two-qubit control operations for circuits composed of parameterized single-qubit rotations and arbitrary entangling gates.

\subsection{Measuring DQFIM on quantum computers using shift-rule and purification}\label{sec:dQFIM_meas_purify}
Now, we propose to measure the DQFIM using its purification and the shift-rule. This method requires preparing the purification of $\rho_L$, and then uses the standard shift-rule to efficiently measure the DQFIM.

It has been shown that the QFIM can be efficiently measured. In particular, one can show that the QFIM is the Hessian of the fidelity~\cite{mari2021estimating,stokes2020quantum} via
\begin{equation}
    \mathcal{F}_{nm}=-\frac{1}{2}\partial_n\partial_m \vert \braket{\chi(\boldsymbol{\theta})}{\chi(\boldsymbol{\theta}_0)}\vert \Big|_{\boldsymbol{\theta}_0=\boldsymbol{\theta}}
\end{equation}
and one can use the shift-rule to efficiently measure the QFIM~\cite{mari2021estimating} given ansatz~\eqref{eq:ansatz_sup}. As we have shown above the DQFIM for data state $\rho_L=L^{-1}\sum_{i} \ket{\psi_i}\bra{\psi_i}$ and unitary $U(\boldsymbol{\theta})$ is equivalent to the QFIM for the purified state
\begin{equation}
    \ket{\chi(\boldsymbol{\theta})}=\frac{1}{\sqrt{L}} \sum_{i=1}^L U(\boldsymbol{\theta})\otimes I_B\ket{\psi_i}\ket{i}
\end{equation} 
where we have
\begin{equation}
\mathcal{Q}_{nm}(\rho_L, U(\boldsymbol{\theta}))=\mathcal{F}_{nm}( \ket{\chi(\boldsymbol{\theta})})\,.
\end{equation} 
Now that we relate the DQFIM to the QFIM of the purification, we can harness known results of efficiently measuring the QFIM. 

Using the shift-rule for the QFIM from Ref.~\cite{mari2021estimating}, we get
\begin{align*}
\mathcal{Q}_{nm}(\boldsymbol{\theta})=-\frac{1}{8}&[\abs{\braket{\chi(\boldsymbol{\theta})}{\chi(\boldsymbol{\theta}+(\boldsymbol{e}_n+\boldsymbol{e}_m)\pi/2)}}^2-\abs{\braket{\chi(\boldsymbol{\theta})}{\chi(\boldsymbol{\theta}+(\boldsymbol{e}_n-\boldsymbol{e}_m)\pi/2)}}^2-\numberthis\label{eq:DQFIM_shift}\\
&\abs{\braket{\chi(\boldsymbol{\theta})}{\chi(\boldsymbol{\theta}+(-\boldsymbol{e}_n+\boldsymbol{e}_m)\pi/2)}}^2+\abs{\braket{\chi(\boldsymbol{\theta})}{\chi(\boldsymbol{\theta}-(\boldsymbol{e}_n+\boldsymbol{e}_m)\pi/2)}}^2]\,,
\end{align*}
where $\boldsymbol{e}_n$ is the basis vector for $n$-th index of parameter vector $\boldsymbol{\theta}$. 
After preparing the purification, one can then efficiently measure the overlaps $\abs{\braket{\chi(\boldsymbol{\theta})}{\chi(\boldsymbol{\theta}+(\boldsymbol{e}_n+\boldsymbol{e}_m)\pi/2)}}^2$ via the inversion test or SWAP test (see Appendix G of Ref.~\cite{haug2021capacity} for a review).

}

\subsection{DQFIM as projection onto training data}

\revB{We have defined the DQFIM as the metric of the parameterized unitary mapped onto the dataset via $U(\boldsymbol{\theta})\rho_L$. 
Note that instead of using $\rho_L$ as a map, one could also consider an alternative definition of the DQFIM via the projector onto the training data
$\Pi_L=\sum_{k=1}^{B_L}\ket{\phi_k}\bra{\phi_k}$ where $\ket{\phi_k}$ are the eigenvectors of $\rho_L=\sum_k p_k\ket{\phi_k}\bra{\phi_k}$ with probabilities $p_k$ and $B_L=\mathrm{rank}(\rho_L)$. 
In particular, this alternative DQFIM could be defined via the projected unitary $U(\boldsymbol{\theta})\Pi_L$, which is an isometry. In fact, in the special case where all training states are orthogonal, this definition is equivalent to the DQFIM as we have $\Pi_L\sim \rho_L$ up to a normalization factor.
Note that we are mainly interested in the rank of the DQFIM $\text{rank}(\mathcal{Q})$, which depends only on the subspace spanned by the training data $\rho_L$, and not the particular distribution of training states itself. Thus, when considering the rank of the DQFIM, there is no difference whether one defines it via the mapped unitary $U(\boldsymbol{\theta})\rho_L$ or a projection $U(\boldsymbol{\theta})\Pi_L$.}

\section{Degrees of freedom of isometries}\label{sec:RLexact}
Now, we calculate the degrees of freedom when learning a training set of $L$ states with a $d$-dimensional unitary 
\begin{equation}\label{eq:U_arb}
U=\sum_{n,k=1}^d u_{nk}\ket{n}\bra{k}
\end{equation}
with $u_{nk}=a_{nk}+ib_{nk}$, where $a_{nk}$, $b_{nk}$ are real parameters. 
First, note that $U$ can be described using $2d^2$ real parameters, however due to $U^\dagger U=I$ and global phase, only $d^2-1$ real parameters are independent. 
We now compute the maximal number of degrees of freedom when $U$ is projected onto a training set of $L$ states. For any set $S_L$ of training states, the rank of $\rho_L$ and equivalently its projector onto the subspace of non-zero eigenvalues $\Pi_L$ is upper bounded by $\text{rank}(\rho_L)=\text{rank}(\Pi_L)\le L$.

We apply $U$~\eqref{eq:U_arb} on a training set $\{\ket{\ell}\}_{\ell=1}^L$ of $L$ states, where $\ket{\ell}$ are computational basis states. 
For a single training state $L=1$, we have $U\ket{1}=\sum_{n=1}^d u_{n1}\ket{n}$. Via training, we can only learn the column vector $u_1=(u_{11},u_{21},\dots,u_{d1})$, which has $2d$ real parameters and $2d-2$ independent parameters due to constraints of global phase and norm $\sum_{n=1}^d\vert u_{n1}\vert^2=1$. However, all other column vectors besides $u_1$ cannot be learned.
With the DQFIM we find $R_1=2d-2$, i.e. $R_1$ indeed counts the number of degrees of freedom that can be learned.
Next, we consider $L=2$. Here, we additional have the state $U\ket{2}=\sum_{n=1}^d u_{n2}\ket{n}$ and we can also learn the vector $u_2=(u_{12},u_{22},\dots,u_{d2})$ with $2d$ real parameters. 
However, due to unitarity, $u_2$ must be orthogonal to $u_1$, which removes two degrees of freedom. Additionally, we have to subtract one parameter for the norm condition. The global phase has already been incorporated in $u_1$, thus $u_2$ holds $2d-3$ degrees of freedom, with the $d\times 2$-dimensional isometry $(u_1,u_2)$ combined having $R_2=4d-5$. 
We can now calculate $R_L$ for any $L$.
For any $2\leq L< d$, adding a training state adds a column $u_L$. $u_L$ has $2d$ real parameters, however it must be orthogonal with all other columns $\ell=1,\dots,L-1$, which reduces the number of independent parameters by $2L-2$. Further, the normalization condition $\vert u_L\vert=1$ removes another free parameter. Thus, the new column $u_L$ has $2d-2L+1$  independent degrees of freedom~\cite{polcari2016representing}. For $L$ states, we have the isometry $U_L=(u_1,\dots, u_L)$, which can be described using 
\begin{equation}\label{eq:analytic}
    R_L=(2d-2)L-(L-1)^2=2dL-L^2-1
\end{equation}
real independent parameters and for $1\leq L\leq d$.
The maximum is reached for $L_\text{c}=d$ with $R_d=d^2-1$ and $U\Pi_d=U$ with $\Pi_d=I$, where we can completely learn $U$. For further increase in $L$ we find that $R_L$ stays constant. As our choice of $U$ is a generic representation of a unitary and our chosen training set has maximal $\text{rank}(\rho_L)=L$, our calculation gives us the upper bound for $R_L$. 
Note that by choosing a more constrained ansatz unitary and training sets $R_L$ can be smaller.

For an arbitrary unitary, the gain in effective dimension by increasing dataset size $L\rightarrow L+1$ is given by $\Delta R_L=R_{L+1}-R_L=\text{max}(2d-2L-1,0)$ for $L\ge1$, and $\Delta R_0=2d-2$ for $L=0$. Thus, the gain decreases with $L$, i.e. with increasing $L$ each additional state reveals less degrees of information of $U$.

\section{Lie-algebra bound on the rank of DQFIM}\label{sec:liebound}

Recall that our ansatz~\eqref{eq:ansatz_sup} consist of $G$ layers with 
\begin{equation}\label{eq:ansatzGen}
    U(\boldsymbol{\theta})=\prod_{k=1}^G U_k(\boldsymbol{\theta}_k)\,\,\text{with}\,\, U^{(k)}(\theta_k)=\prod_{n=1}^K\exp(-i\theta_{kn} H_n)
\end{equation}
where $U^{(k)}(\theta_k)$ is the unitary of the $k$th layer. Here, $H_n$ are $K$ hermitian matrices and $\boldsymbol{\theta}_k=\{\theta_{k1},\dots,\theta_{kn}\}$ are the parameters of the $k$th layer. The total parameter vector $\theta=\{\boldsymbol{\theta}_1,\dots,\boldsymbol{\theta}_G\}$ of the ansatz has $M=GK$ parameters. 

To simplify the notation, we treat each parameter entry as its own layer and relabel each generators $H_k$ such that we can write the ansatz as
\begin{equation}\label{eq:ansatzdla}
U(\boldsymbol{\theta})=\prod_{k=1}^M\exp(-i\theta_{k} H_k)\,.
\end{equation}
First, we define the generators of the ansatz $U$~\cite{d2021introduction,larocca2021theory}:
\begin{defn}[Set of generators $\mathcal{T}$]\label{def:generators}
Consider ansatz~\eqref{eq:ansatzdla}. The set of generators $\mathcal{T}=\{H_k\}_{k=1}^K$ (with size $|\mathcal{T}|=K$) are defined as the set of Hermitian operators that generate the unitaries of each layer of $U(\boldsymbol{\theta})$.
\end{defn}
Further the dynamical Lie Algebra $\mathfrak{g}$ is given by:
\begin{defn}[Dynamical Lie Algebra (DLA)]\label{def:dynamical_lie_algebra-SM}
Consider the generators $\mathcal{T}$ according to Def.~\ref{def:generators}. The DLA $\mathfrak{g}$ is generated by repeated nested commutators of the operators in $\mathcal{T}$
\begin{equation}
\mathfrak{g}=\mathrm{span}\left\langle iH_1, \ldots, iH_K \right\rangle_\text{Lie}\,,
\end{equation}
where $\left\langle \mathcal{S}\right\rangle_\text{Lie}$ is the Lie closure, which is the set obtained by repeatedly taking the commutator of the elements in $\mathcal{S}$. 
\end{defn}

Next, we show that the DLA bounds the rank of the DQFIM.
First, lets recall the entries of the matrix of the DQFIM
\begin{align*}
&\mathcal{Q}_{nm}(S_L,U(\boldsymbol{\theta}))=4\text{Re}(\text{tr}(\partial_n U \rho_L \partial_m U^\dagger)-\text{tr}(\partial_n U\rho_L U^\dagger)\text{tr}(U \rho_L\partial_m U)^\dagger)\,.\numberthis\label{eq:Fisher_supdla}
\end{align*}
where we shorten $U=U(\boldsymbol{\theta})$ and $\partial_n U$ the derivative in respect to the $n$th element of parameter vector $\boldsymbol{\theta}$.

We now restate Theorem~1 of the main text for convenience.
\begin{theorem}\label{th:liebound_sup}
    The maximal rank $R_L$ of the DQFIM  is upper bounded by the dimension of the dynamical Lie algebra (DLA) $\dim(\mathfrak{g})$
    \begin{equation}
    R_L\leq\dim(\mathfrak{g})\,,
    \end{equation}
    where $\mathfrak{g}=\mathrm{span}\left\langle iH_1, \ldots, iH_K \right\rangle_\text{Lie}$ is generated by the repeated nested commutators of the generators $H_k$ of the unitary $U(\boldsymbol{\theta})$.
\end{theorem}
The proof follows in analogy to the bound of the QFIM (i.e. $R_1$) of Ref.~\cite{larocca2021theory}.

First, we note that $R_L\le R_\infty$ as $L$ simply increases the dimension spanned by the dataset $\rho_L$. Now, we assume $S_L$ spans the complete relevant Hilbertspace with $\rho_L=I/d$ and thus we can simplify to the unitary QFIM
\begin{align*}
&\mathcal{Q}_{nm}=4\text{Re}(d^{-1}\text{tr}(\partial_n U \partial_m U^\dagger)-d^{-2}\text{tr}(\partial_n U U^\dagger)\text{tr}(U\partial_m U^\dagger))\,.\numberthis\label{eq:Fisher_supdlaUp}
\end{align*}
As we have $\partial_n \exp(-i\theta_n H_n)= -iH_n\exp(-i\theta_n H_n)$, we can write the derivatives as 
\begin{equation}\label{eq:deriv}
\partial_n U(\boldsymbol{\theta})=-iU_{n+1\rightarrow M}H_nU_{1\rightarrow n}\,.
\end{equation}
Here, we define  
\begin{equation}
U_{m\rightarrow n}=U_n U_{n-1} \dots U_{m+1}U_m
\end{equation}
as the propagator from layer $m$ to layer $n$ and
\begin{equation}
    \tilde{H}_k=U_{1\rightarrow k}^\dagger H_k U_{1\rightarrow k}\,.
\end{equation}

Using above expressions, we find for the first term of the DQFIM~\eqref{eq:Fisher_supdlaUp} 
\begin{align*}
    &\text{Re}(\text{tr}(\partial_n U \partial_m U^\dagger)) =
    \text{Re}( i(-i)\text{tr}(U_{n+1\rightarrow M}  H_n U_{1\rightarrow n} U_{1\rightarrow m}^\dagger H_m  U_{m+1\rightarrow M}^\dagger))=
    \text{Re}(\text{tr}(\tilde{H}_n\tilde{H}_m))\,.
 \end{align*}
Similarly, we find for the second term of~\eqref{eq:Fisher_supdlaUp} 
\begin{align*}
    &\text{Re}(\text{tr}(\partial_n U U^\dagger)\text{tr}(U \partial_m U^\dagger))=
    \text{Re}(i(-i)\text{tr}(U_{n+1\rightarrow M} H_n U_{1\rightarrow n} U_{1\rightarrow M}^\dagger)\text{tr}(U_{1\rightarrow M} U_{1\rightarrow m}^\dagger  H_m U_{m+1\rightarrow M}^\dagger))=
     \text{Re}(\text{tr}(\tilde{H}_n)\text{tr}(\tilde{H}_m))\,.
\end{align*}

We combine these results to get the DQFIM as
\begin{equation}\label{eq:tildeQFIM}
\begin{split}
   \mathcal{Q}_{mn}^I=\text{Re}(d^{-1}\text{tr}(\tilde{H}_n\tilde{H}_m)-d^{-2}\text{tr}(\tilde{H}_n)\text{tr}(\tilde{H}_m))
\end{split}
\end{equation}
Note that $H_k$ are elements of the DLA $\mathfrak{g}$. As the unitaries $U_n$ of the ansatz are also elements of the dynamical Lie group generated by $\mathfrak{g}$, a product of $H_k$ with any $U_n$ will yield another element of the dynamical Lie group.
Thus, we can always expand $\tilde{H}_k$ using the DLA as a basis with $\text{dim}(\mathfrak{g})$ elements:
\begin{equation}
    \tilde{H_k} = \sum_{m=1}^{\text{dim}(\mathfrak{g})}  a_{m}^{(k)} \chi_m\,,
\end{equation}
where $a_{m}^{(k)}$ are real coefficients and $\{\chi_m\}_{m=1}^{\text{dim}(\mathfrak{g})}$ are a basis of the DLA $\mathfrak{g}$. 
Thus, the matrix of the DQFIM $\mathcal{Q}$ can be expressed in a basis with $\text{dim}(\mathfrak{g})$ elements. Thus, the rank of $\mathcal{Q}$ is upper bounded by the dimension of the DLA $\mathfrak{g}$
\begin{equation}
    R_L\le\text{rank}(\mathcal{Q}^I)\le \text{dim}(\mathfrak{g})\,.
\end{equation}

\section{Ansatz unitaries}\label{sec:ansatz}
\begin{figure*}[htbp]
	\centering	
\subfigimg[width=0.7\textwidth]{}{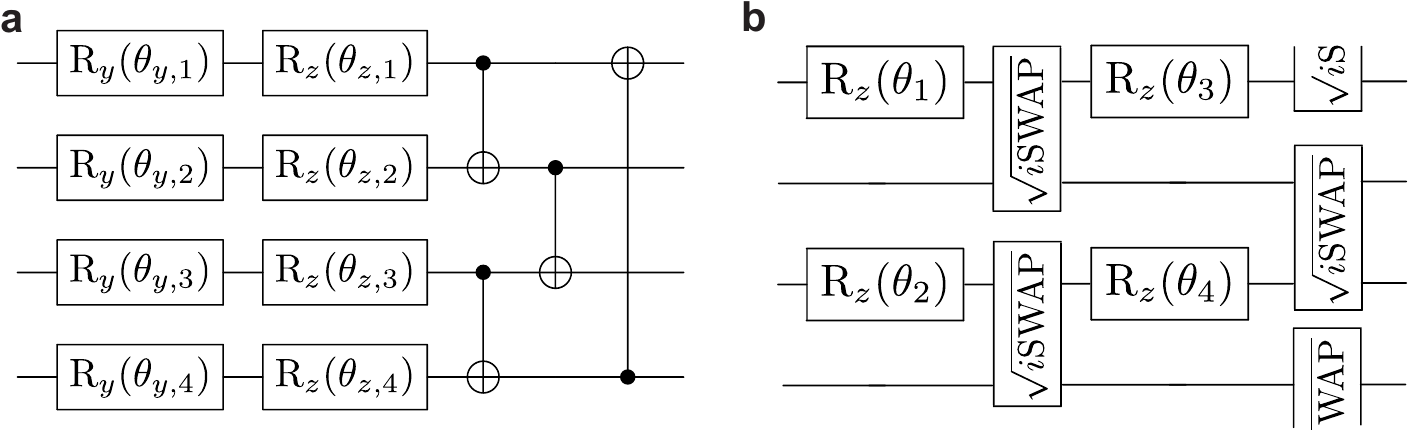}\hfill
	\caption{ Ansatz unitaries for the main text. The circuits are repeated for $G$ layers. 
 \idg{a} Hardware-efficient ansatz $U_\text{HE}$ consisting of parameterized $y$, $z$ rotations and CNOT gates. Has no symmetries and can realize arbitrary $N$-qubit unitaries for sufficient depth.
 \idg{b} $U_{XY}$ circuit inspired by $XY$-model~\eqref{eq:XY}. Composed of single qubit $z$ rotations and nearest-neighbor $\sqrt{\text{iSWAP}}$ gates, arranged with periodic boundary condition. Commutes with particle number operator $P$.
	}
	\label{fig:ansatze}
\end{figure*}

The ansatz unitaries used in the maint text are shown in Fig.~\ref{fig:ansatze}.
We assume that the considered unitaries have a periodic structure of $G$ layers with 
\begin{align*}
    &U(\boldsymbol{\theta})=\prod_{k=1}^G U_k(\boldsymbol{\theta}_k) \,,\,\,U^{(k)}(\theta_k)=\prod_{n=1}^K\exp(-i\theta_{kn} H_n)\label{eq:ansatz_sup}\numberthis
\end{align*} 
where $U^{(k)}(\theta_k)$ is the unitary of the $k$th layer. Here, $H_n$ are $K$ hermitian matrices and $\boldsymbol{\theta}_k=\{\theta_{k1},\dots,\theta_{kn}\}$ are the parameters of the $k$th layer. The total parameter vector $\theta=\{\boldsymbol{\theta}_1,\dots,\boldsymbol{\theta}_G\}$ of the ansatz has $M=GK$ parameters.

In Fig.\ref{fig:ansatze}a, we show a hard-ware efficient ansatz $U_\text{HE}$, which produces highly random circuits which span the full Hilbertspace nearly uniformly and are known to be hard to simulate classically. Overparameterization requires for this circuit exponentially many parameters.

In Fig.\ref{fig:ansatze}b, we show an ansatz $U_\text{XY}$ that overparameterizes in polynomial depth.
This ansatz is inspired from the integrable XY Hamiltonian with random field $h_k$
\begin{equation}\label{eq:XY}
H_{XY}=\sum_{k=1}^N h_k \sigma_k^z +\sum_{k=1}^N( \sigma_k^x\sigma_{k+1}^x+\sigma_k^y\sigma_{k+1}^y)
\end{equation}
$H_{XY}$ commutes with the particle number operator
$P=\sum_{k=1}^N\frac{1}{2}(1-\sigma^z_k)$
where $\sigma^z_k$ is the Pauli $z$ operator acting on qubit $k$. In particular, we have $[H_{XY},P]$.
Its time evolution $U=\exp(-iH_{XY}t)$ also conserves the symmetry, i.e. $[U,P]=0$. 

The ansatz $U_{XY}$ shown in Fig.~\ref{fig:ansatze}b can represent the time evolution of the Hamiltonian. $U_{XY}$ consists of parameterized $z$-rotations and nearest-neighbor $\sqrt{\text{iSWAP}}=\exp(i\pi/8(\sigma_k^x\sigma_{k+1}^x+\sigma_k^y\sigma_{k+1}^y))$ gates. The ansatz conserves particle number symmetry as well, i.e. $[U_{XY},P]=0$ One can think of this model similar to a Trotterized version of the time evolution $U=\exp(-iH_{XY}t)$.  The generators of $U_{XY}$ are the Pauli operators of $H_{XY}$. Thus, the time-evolution operator $U_{XY}$ spans the same dynamical Lie algebra as the time evolution generated by $H_\text{XY}$ and can represent any time evolution of $H_\text{XY}$~\cite{kokcu2022algebraic}. The dimension of the dynamical Lie-algebra spanned by  $U_{XY}$ scales polynomial with qubit number $N$ and thus can be overparameterized with polynomially many parameters $M$~\cite{larocca2021theory}. 
For random product states $W_\text{prod}$ as training set, we we find via numerical extrapolation $R_1=2N^2-3N+2$ and $R_\infty=2N^2-1$.
As another training set, we choose $W_{p=1}$ with $P=1$ particles, which consists of arbitrary superpositions of permutations of the basis states $\ket{10\dots0}$. We have $P\ket{\psi_\ell}=\ket{\psi_\ell}$ for any state $\ket{\psi_\ell}\in W_{p=1}$. These states live in an effective $N$-dimensional subspace, yielding $R_1=2N-2$ and $R_\infty=N^2-1$.

\section{Additional results on XY circuit learning}\label{sec:trainXY}
We now show additional numerical results on training with the $U_{XY}$ ansatz which conserves particle number.

\revA{
In Fig.~\ref{fig:traintestXY}, we study overparameterization and generalization as function of $M$.  We study test and training error for the $U_{XY}$ ansatz for training with product state ensemble $W_{\mathrm{prod}}$. Using the analytic rank of DQFIM of the main text, we find $L_\text{c}\approx 2$.
For sufficient $M$, we have $C_\text{train}\approx 0$ for $L=1$ and $L=2$ training data. Generalization with $C_\text{test}\approx 0$ requires $L\ge2$ when testing in-distribution $W_{\mathrm{prod}}$ or out-of-distribution $W_{p=1}$, matching the result predicted using the DQFIM.

\begin{figure*}[htbp]
	\centering	
 \subfigimg[width=0.28\textwidth]{}{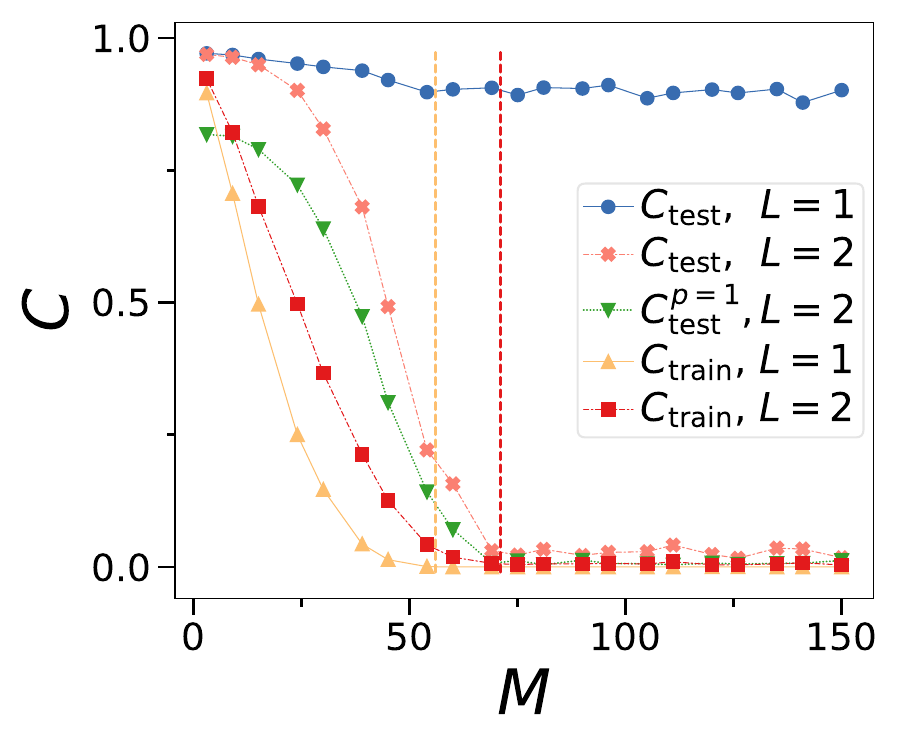}
	\caption{Unitary learning with $U_{XY}$ ansatz. We show $C_\text{test}$ and $C_\text{train}$ against $M$ with $U_{XY}$ ansatz for $N=6$ qubits, where we train with $L=1$ and $L=2$ product states $W_{\mathrm{prod}}$. We use $W_{\mathrm{prod}}$ as test states, except for green dotted curve $C_\text{test}^{p=1}$ where we show out-of-distribution generalization with symmetric test states $W_{p=1}$. The dashed vertical lines indicate $R_1$ (yellow) and $R_2$ (red).
	}
	\label{fig:traintestXY}
\end{figure*}

Now, we use training data which respects particle number $P=\sum_{k=1}^N\frac{1}{2}(1-\sigma^z_k)$, i.e. the training states are sampled from from $W_{p=1}$ which are states with Hamming weight $1$. Using the rank of the DQFIM, we compute $L_\text{c}\approx 2 R_\infty/R_1\approx N$.

In Fig.~\ref{fig:singleN}a we study $C_\text{test}$ against $L$, $M$ for $U_{XY}$ when training with particle-conserved ensemble $W_{p=1}$. Generalization improves with $M$ and $L$, where the lower bound $C_\text{test}\sim 1-(L/L_\text{c})^2$~\cite{sharma2022reformulation} is saturated for $M\ge M_\text{c}$.
In Fig.~\ref{fig:singleN}b, we study generalization in the overparameterized regime. We find $C_\text{test}\sim 1-(L/L_\text{c})^2$ independent of $N$ , which matches the theoretical result of Ref.~\cite{sharma2022reformulation}.
 In Fig.~\ref{fig:singleN}c, we find that for overparameterized models $E\sim N^2$ with a clear peak at $L\approx L_\text{c}$. 
In Fig.~\ref{fig:singleN}d we study 
the steps $E$ needed to converge against $M$ for different $N$ for overcomplete data $L\gg L_\text{c}$. We observe $E\sim N^2$. At $M_\text{c}$ the steps needed to converge sharply decreases, indicating the transition to an  optimization landscape where the global minimum can be reached easily.
\begin{figure*}[htbp]
	\centering	
 \subfigimg[width=0.28\textwidth]{a}{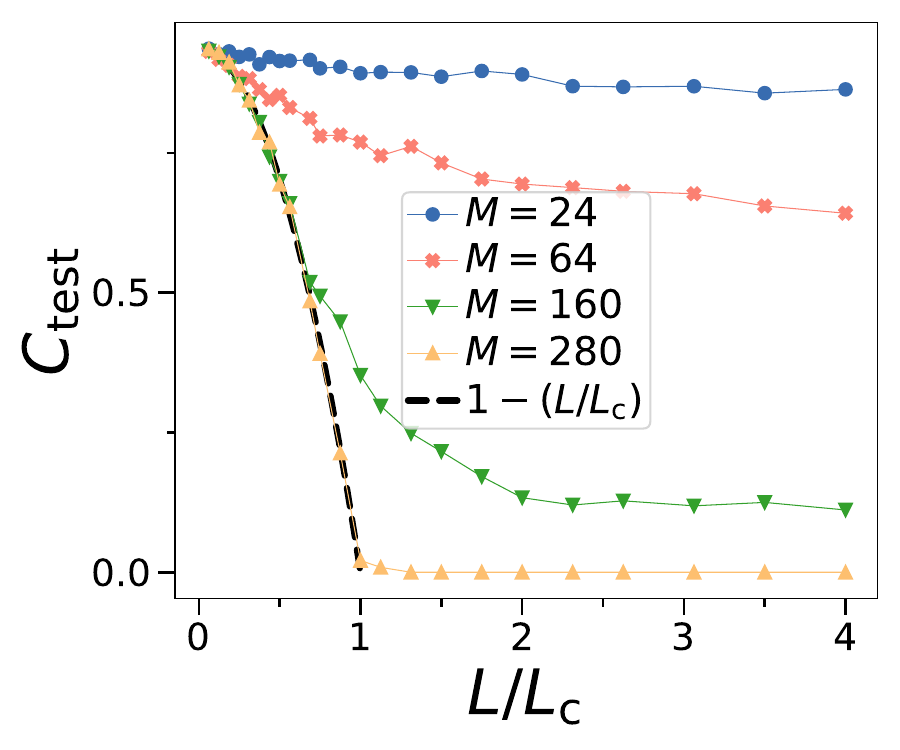}
\subfigimg[width=0.28\textwidth]{b}{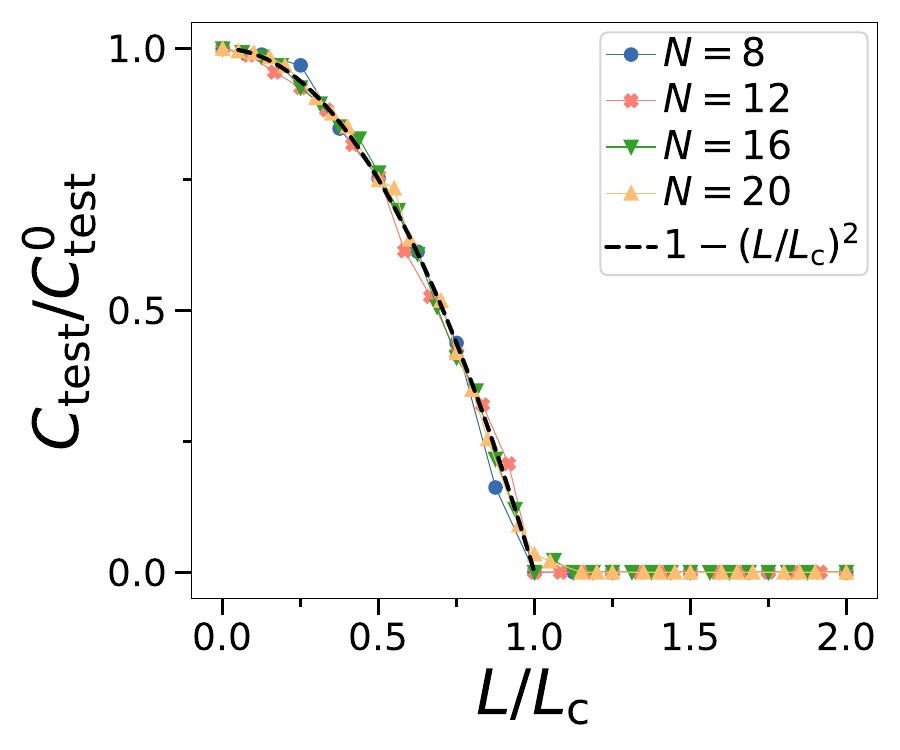}
\subfigimg[width=0.28\textwidth]{c}{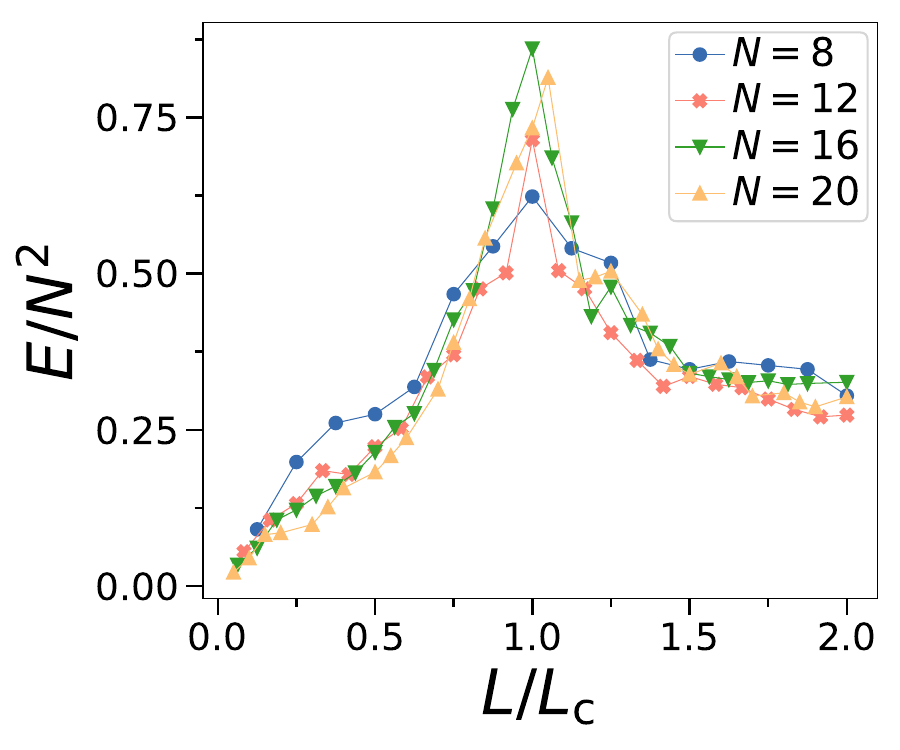}
\subfigimg[width=0.28\textwidth]{d}{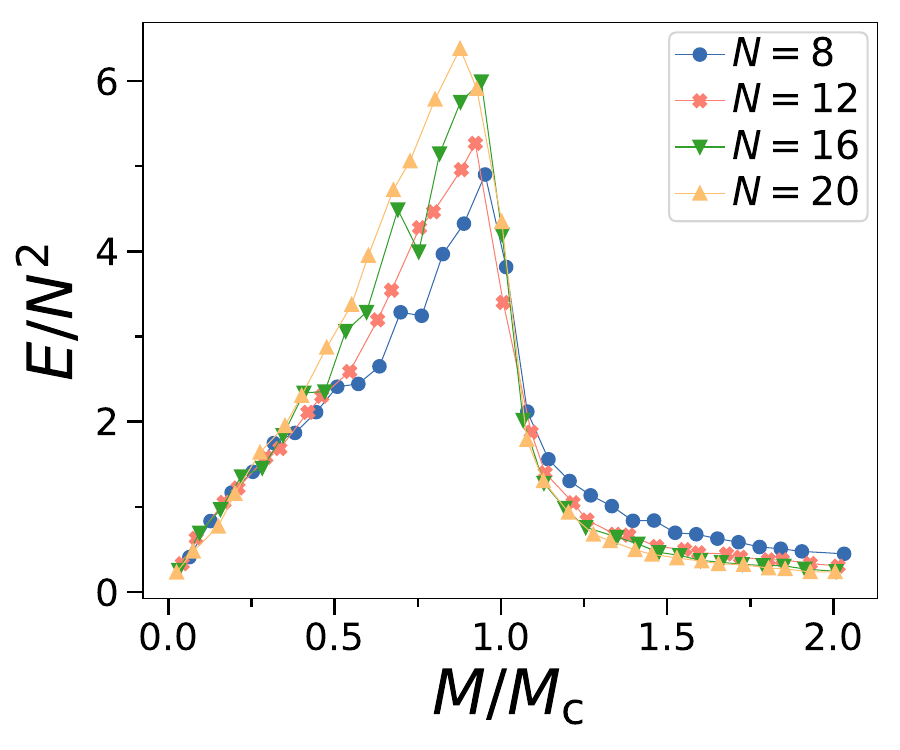}
	\caption{Unitary learning of particle number conserving ansatz $U_{XY}(\boldsymbol{\theta})$ with symmetric states $\ket{\psi_\ell}\in W_{p=1}$.
\idg{a} $C_\text{test}$ against $L$ for different $M$ with $N=16$, $L_\text{c}=N$. Black dashed  line is $C_\text{test}\sim 1- (L/L_\text{c})^2$.
 \idg{b}~Test error $C_\text{test}$ relative to error without training $C_\text{test}^0$ for varying training set size $L$ and $M\gg M_\text{c}$ where we use symmetric data $\ket{\psi_\ell}\in W_{p=1}$. We find $C_\text{test}/C_\text{test}^0=1- (L/L_\text{c})^2$, where $L_\text{c}=N$ and we average over 10 random instances.
  \idg{c} Training steps $E$ needed to converge to $C_\text{train}<10^{-4}$ against $L$ for $M\gg M_\text{c}$.
 \idg{d}~Training steps $E$ required to find $C_\text{test}<10^{-3}$. 
	}
	\label{fig:singleN}
\end{figure*}

}

\begin{figure*}[htbp]
	\centering	
\subfigimg[width=0.24\textwidth]{a}{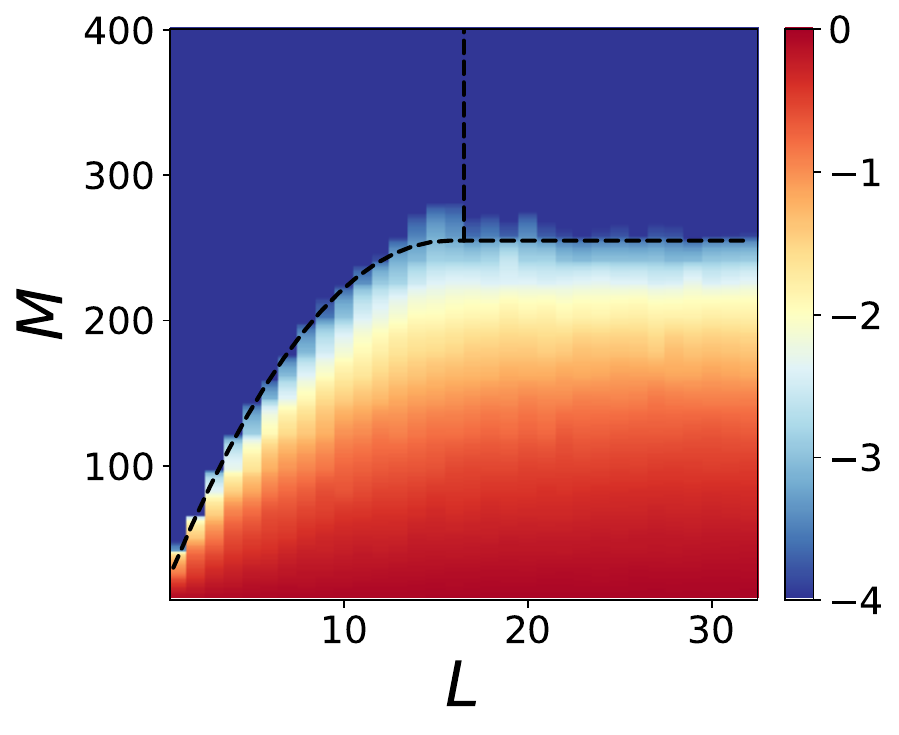}
\subfigimg[width=0.24\textwidth]{b}{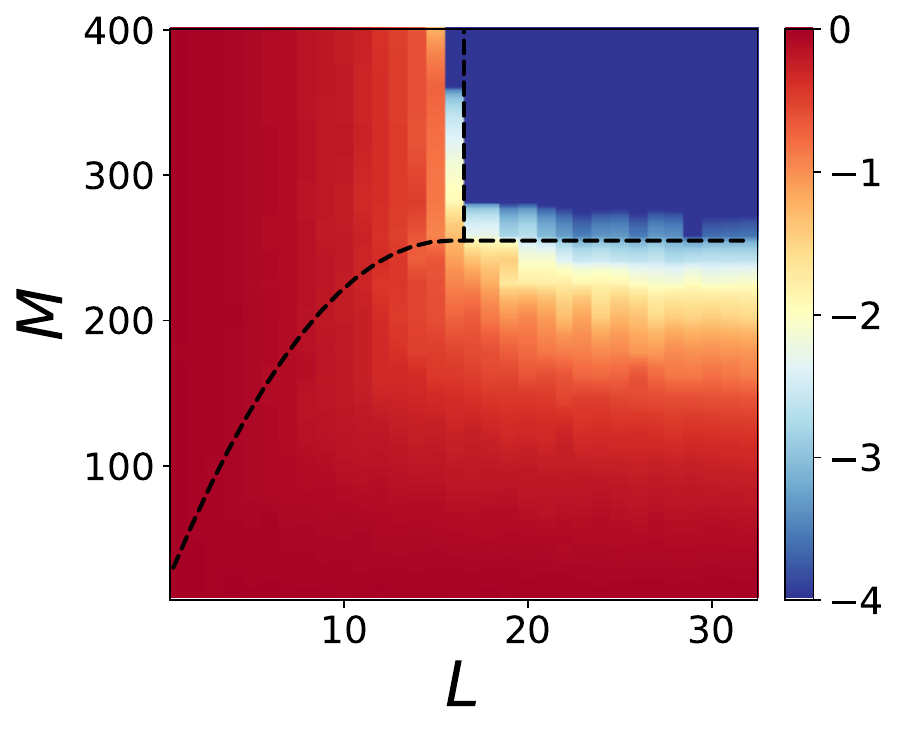}
\subfigimg[width=0.24\textwidth]{c}{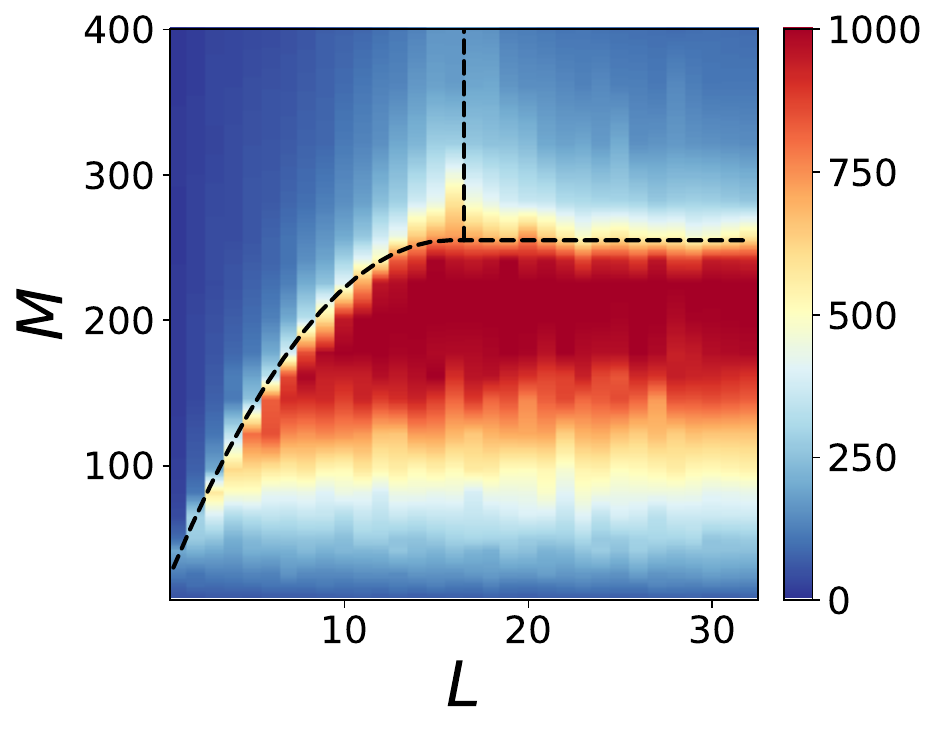}
	\caption{Median error for training $U_{XY}$ for symmetric data $W_{p=1}$ for $N=16$ qubits.
 \idg{a} $C_\text{train}$ against $M$ and $L$. Color shows logarithm $\log_{10}(C_\text{train})$.
 \idg{b} $C_\text{test}$ against $M$ and $L$.
 \idg{c} Number of training steps $E$ until reaching $C_\text{train}< 10^{-3}$.
	}
	\label{fig:single2d}
\end{figure*}

Fig.~\ref{fig:single2d}, we show $C_\text{train}$, $C_\text{test}$ and number of iterations $E$ against $L$ and $M$ for symmetric data $W_{p=1}$. We find training and test error matches closely the transitions derived from $R_L$ which are shown as black dashed lines.

\begin{figure*}[htbp]
	\centering	
\subfigimg[width=0.24\textwidth]{a}{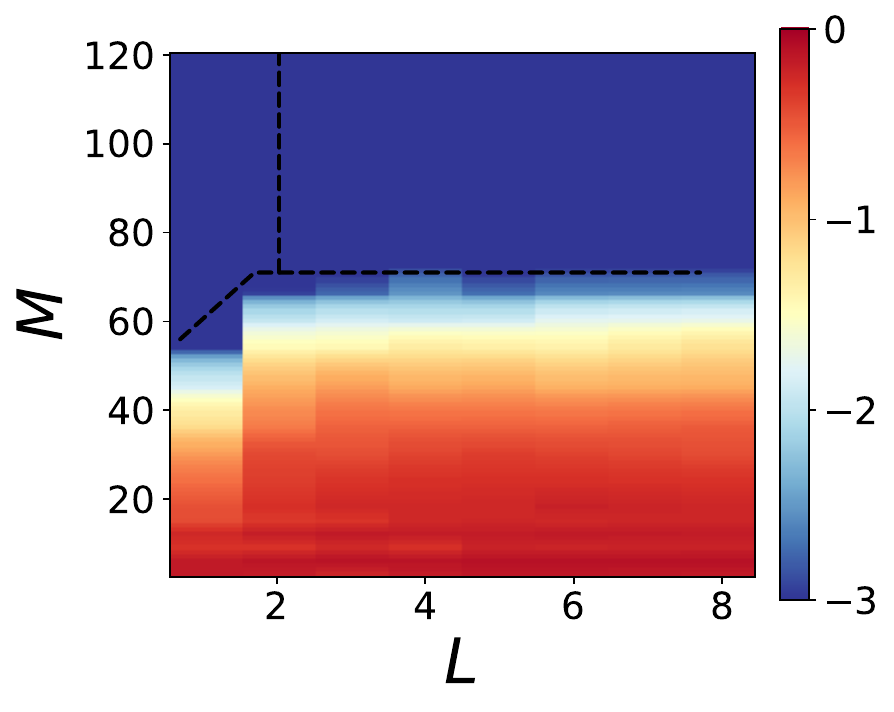}
\subfigimg[width=0.24\textwidth]{b}{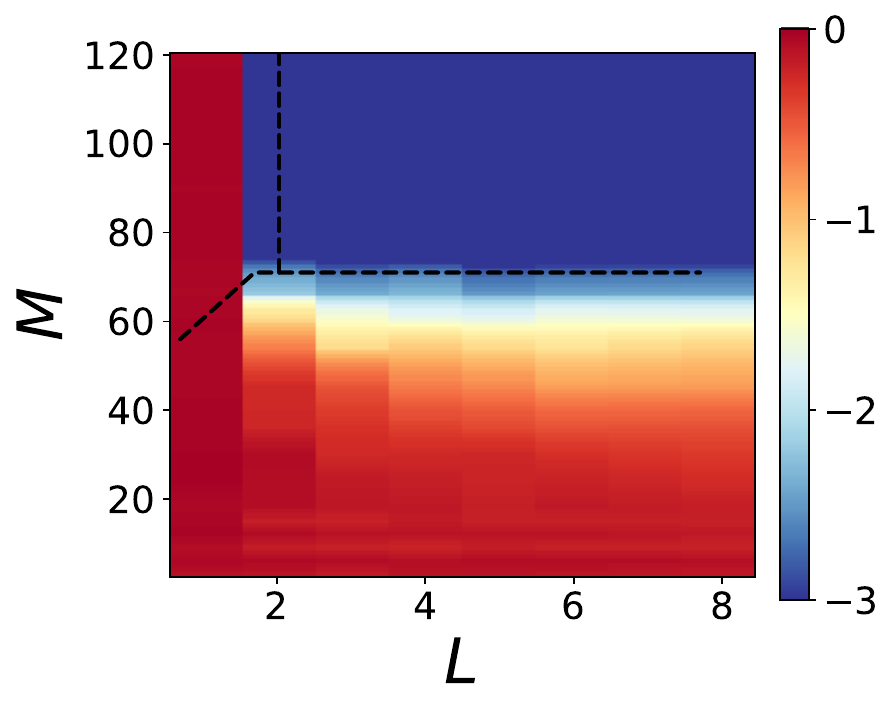}
\subfigimg[width=0.24\textwidth]{c}{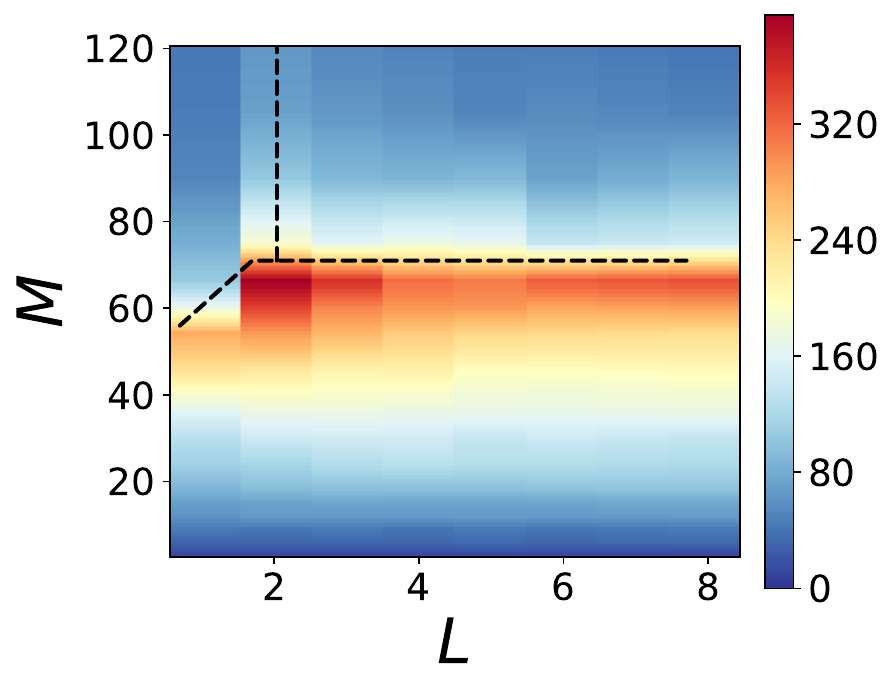}
\subfigimg[width=0.24\textwidth]{d}{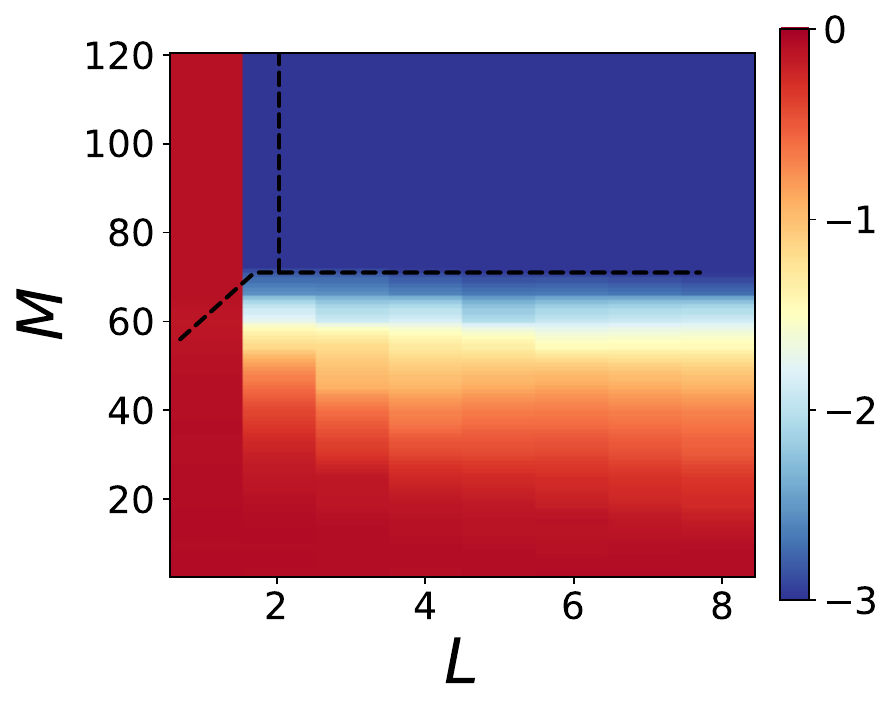}
	\caption{Median error for training $U_{XY}$ ansatz for random product state data sampled from $W_\text{prod}$ and $N=6$ qubits.
 \idg{a} $C_\text{train}$ against $M$ and $L$. Color shows logarithm $\log_{10}(C_\text{train})$.
 \idg{b} $C_\text{test}$ against $M$ and $L$ tested against $W_\text{prod}$.
 \idg{c} Number of training steps $E$ until reaching $C_\text{train}< 10^{-3}$.
  \idg{d} Out-of-distribution generalization $C_\text{test}$ against $M$ and $L$, where we trained with $W_\text{prod}$, but tested with $W_{p=1}$.
	}
	\label{fig:XYprod2d}
\end{figure*}

Next, in Fig.\ref{fig:XYprod2d}a-c we study $C_\text{train}$ and $C_\text{test}$ for the $U_{XY}$ ansatz. Here,  we train using random product states $W_{\text{prod}}$ which do not respect particle number symmetry. Here, we generalize already for $L\ge2$.

Next, in Fig.\ref{fig:XYprod2d}d we study out-of-distribution generalization with the $U_{\text{XY}}$ ansatz. 
Here, the training states are drawn from set of product states $W_{\text{prod}}$ which break particle number symmetry $P=\sum_{k=1}^N\frac{1}{2}(1-\sigma^z_k)$, i.e. $P\ket{\psi}$ . However, the test test data is drawn from a different distribution $W_{p=1}$ (states with Hamming weight 1) which conserve particle number symmetry $P$. We find that although we train with a different distribution that we use for testing, we achieve the same test error as when training directly with $W_{\text{prod}}$. In particular, we generalize for $L\ge2$. 

In contrast, for in-distribution generalization where one both trains and test with states from $W_{p=1}$ (which respect particle number symmetry), we achieve generalization only for $L\ge N$, where $N$ is the number of qubits. Refer to Fig.~\ref{fig:singleN}a of $C_\text{test}$ against $L$.

\section{Generalization with further circuit models}\label{sec:training_gen}
\begin{figure*}[htbp]
	\centering	
\subfigimg[width=0.9\textwidth]{}{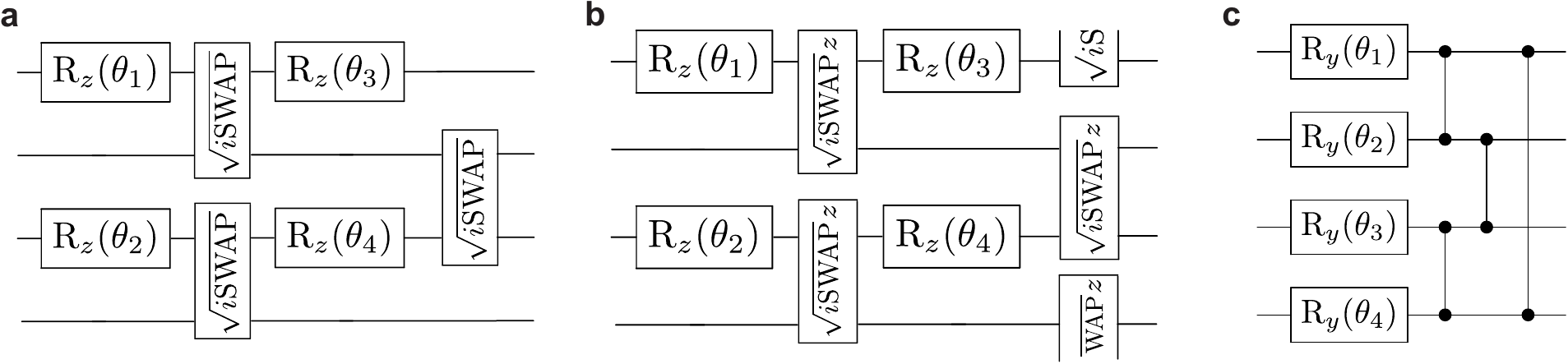}\hfill
	\caption{ Ansatz unitaries  for the the supplemental materials. The circuits are repeated for $G$ layers. 
 \idg{a} $U_{XY}^{\text{open}}$ circuit with open boundary condition, i.e. the $\sqrt{\text{iSWAP}}$ do not cross from the first to the last qubit in contrast to $U_{XY}$. Commutes with particle number operator $P$ and for sufficient depth can realize any time evolution generated by $\exp(-iH_{XY}^{\text{open}} t)$.
 \idg{b} $U_{XXZ}$ circuit related to evolution of Heisenberg model $H_{XXZ}$.
 Composed of parameterized single qubit $z$ rotations and the $\sqrt{\text{iSWAP}}z$ gate defined in the text.
 \idg{c} Real-valued ansatz $U_\text{Y-CZ}$ consisting of $y$-rotations and control-$Z$ gates in a nearest-neighbor chain configuration. 
	}
	\label{fig:ansatze_sup}
\end{figure*}

We now study further circuit models which were used to study generalization in other publications.
In Fig.\ref{fig:ansatze_sup} we show three ansatz unitaries which we then proceed to study.

In Fig.\ref{fig:ansatze_sup}a we show the $U_{XY}^\text{open}$ circuit, which is the same as the $U_{XY}$ circuit but with open boundary conditions, i.e. the $\sqrt{\text{iSWAP}}$ gates that interact between the first and last qubit are removed. This ansatz conserves particle number $P$.

In Fig.\ref{fig:ansatze_sup}b we show the $U_{XXZ}$ ansatz, which is composed of parameterized $z$ rotations and the $\sqrt{\text{iSWAP}}z=\sqrt{\text{CZ}}\sqrt{\text{iSWAP}}$ gate, where $\sqrt{\text{CZ}}=\text{diag}(1,1,1,i)$ is the square-root of the control-$Z$ gate. This ansatz conserves particle number $P$.

In Fig.\ref{fig:ansatze_sup}c we show the $U_{Y-CZ}$ ansatz~\cite{haug2021capacity}, consisting of parameterized $y$-rotations and control-$Z$ gate $\text{diag}(1,1,1,-1)$. Due to its connection to Cluster-state generation, it overparameterizes with a polynomial number of parameters with $R_L\propto N^2$.

\begin{figure*}[htbp]
	\centering	
\subfigimg[width=0.28\textwidth]{a}{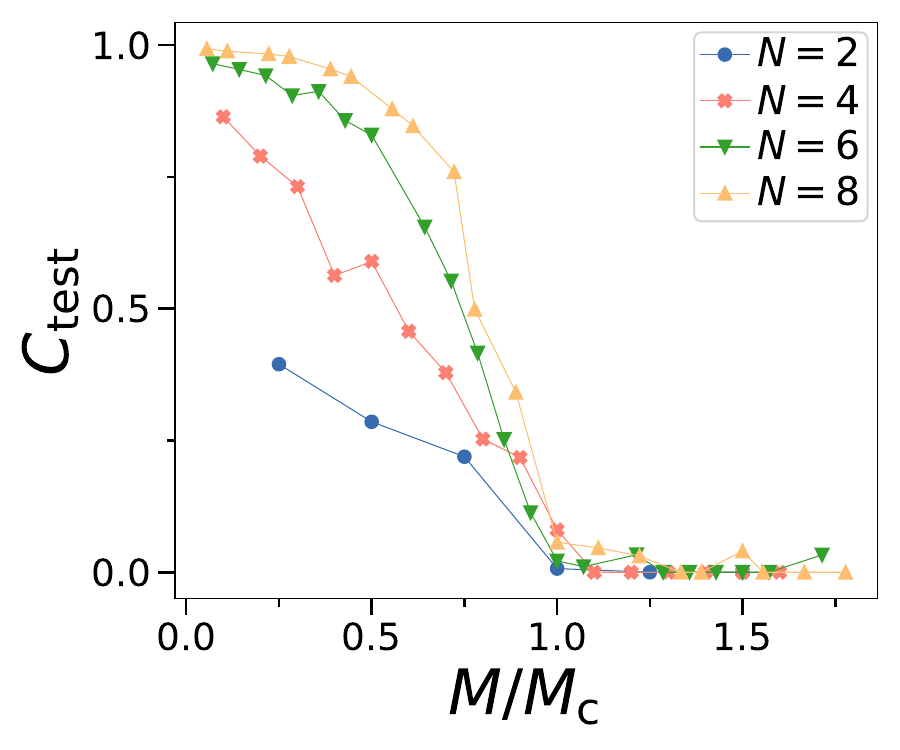}
\subfigimg[width=0.28\textwidth]{b}{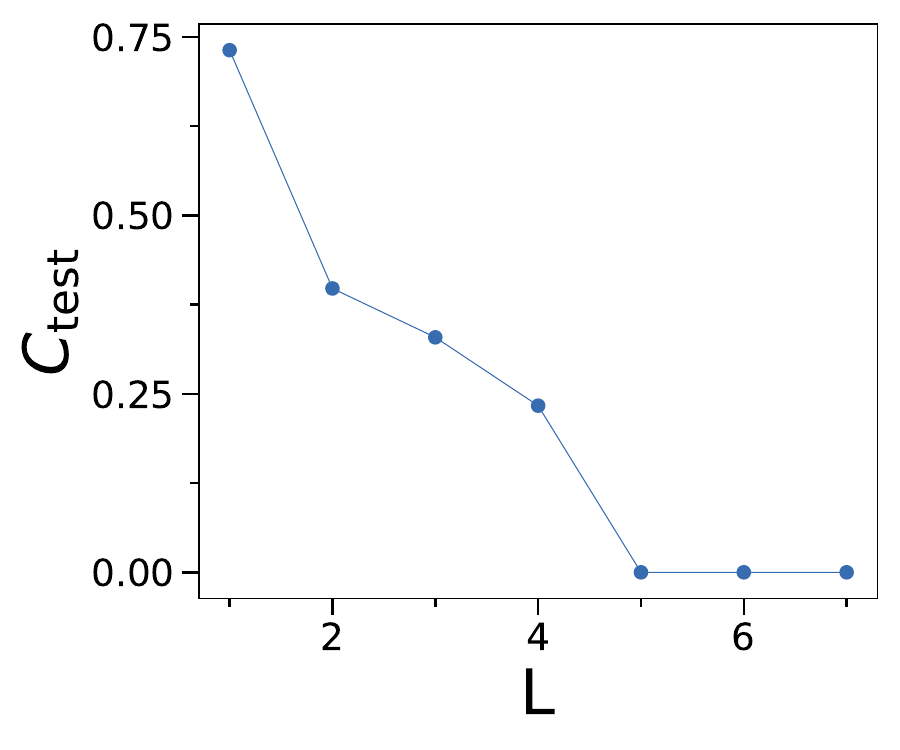}
	\caption{\idg{a} Test error $C_\text{test}$ for $U_{XY}^\text{open}$ ansatz and for $L=1$ product training states against circuit parameters $M$. $C_\text{test}$ is averaged over 20 random instances. $M_\text{c}$ is determined with the DQFIM. 
 \idg{b}  $C_\text{test}$ for $U_{XZZ}$ ansatz for product training states against $L$ for $N=4$ and $M=100$.  
	}
	\label{fig:test_models}
\end{figure*}

Here, we study the number of training states needed for generalization for the above describe ansatze for the unitary learning task.
First, in Fig.\ref{fig:test_models}a we study the $U_{XY}^\text{open}$ ansatz shown in Fig.\ref{fig:ansatze_sup}a. This ansatz describes the evolution of the $H_{XY}^\text{open}=\sum_{k=1}^N h_k \sigma_k^z +\sum_{k=1}^{N-1}( \sigma_k^x\sigma_{k+1}^x+\sigma_k^y\sigma_{k+1}^y)$ Hamiltonian with open boundary conditions. The difference to $H_{XY}$ is the absence of interaction between first and last qubit. We find generalization for $L=1$ training states and $M\ge M_\text{c}$ when using random product states as training data.
A similar ansatz was studied numerically in Ref.~\cite{gibbs2022dynamical}. It was shown numerically that only $1$ training state was needed for generalization, and $O(N^2)$ gates for successful training. 
Here, we explain this result with the DQFIM. In particular, our ansatz has the maximal rank of the DQFIM with $R_L=R_1=R_\infty=N^2$ for all $N$. This implies that $L_\text{c}=1$ training state is sufficient to get an overcomplete model and achieve generalization.

Next, we study the $U_{XXZ}$ ansatz shown in Fig.\ref{fig:ansatze_sup}b. This model can describes the evolution of the $H_{XXZ}=\sum_{k=1}^N h_k \sigma_k^z +\sum_{k=1}^N( \sigma_k^x\sigma_{k+1}^x+\sigma_k^y\sigma_{k+1}^y+\Delta \sigma_k^z\sigma_{k+1}^z)$ Hamiltonian. A similar ansatz was studied in Ref.~\cite{gibbs2022dynamical}. It was numerically shown that $5$ training states are needed for generalization for $N=4$. In Fig.\ref{fig:test_models}b, we show the test error of the $U_{XXZ}$ ansatz against $L$ in the overparameterized regime and indeed find the test error vanishes for $L\ge5$.
Using the DQFIM, we find $R_1=24$ and $R_\infty=R_{L_\text{c}}=51$ with  $L_\text{c}=5$, matching the numerical results. 
Thus, the DQFIM accurately predicts the needed training states.
Note that the approximation $L_\text{c}\approx 2R_\infty/R_1=4.25$ gives a good estimation of the number of needed training states as well.

\begin{figure*}[htbp]
	\centering	
\subfigimg[width=0.24\textwidth]{a}{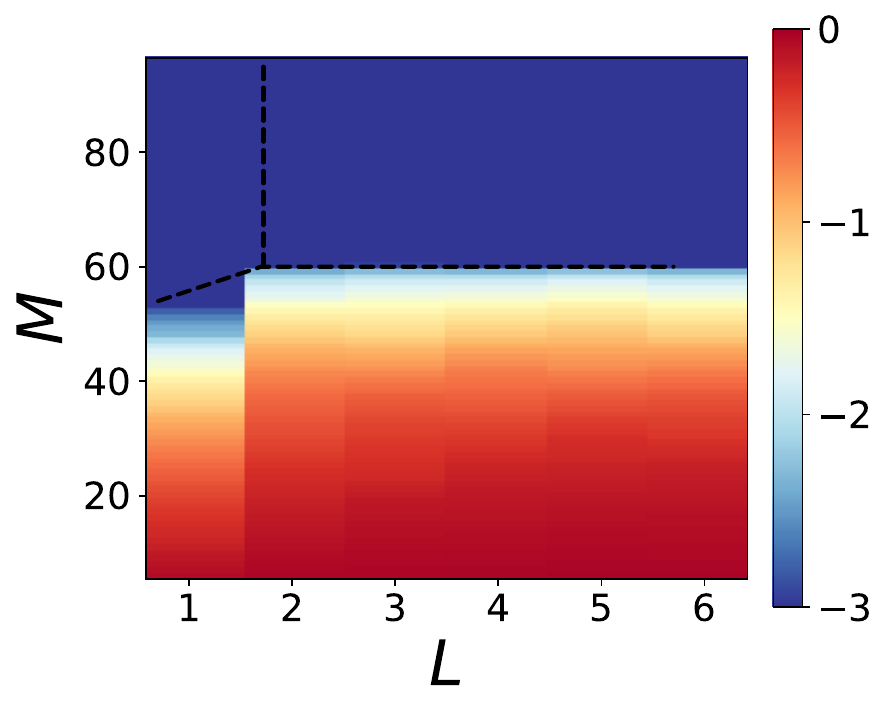}
\subfigimg[width=0.24\textwidth]{b}{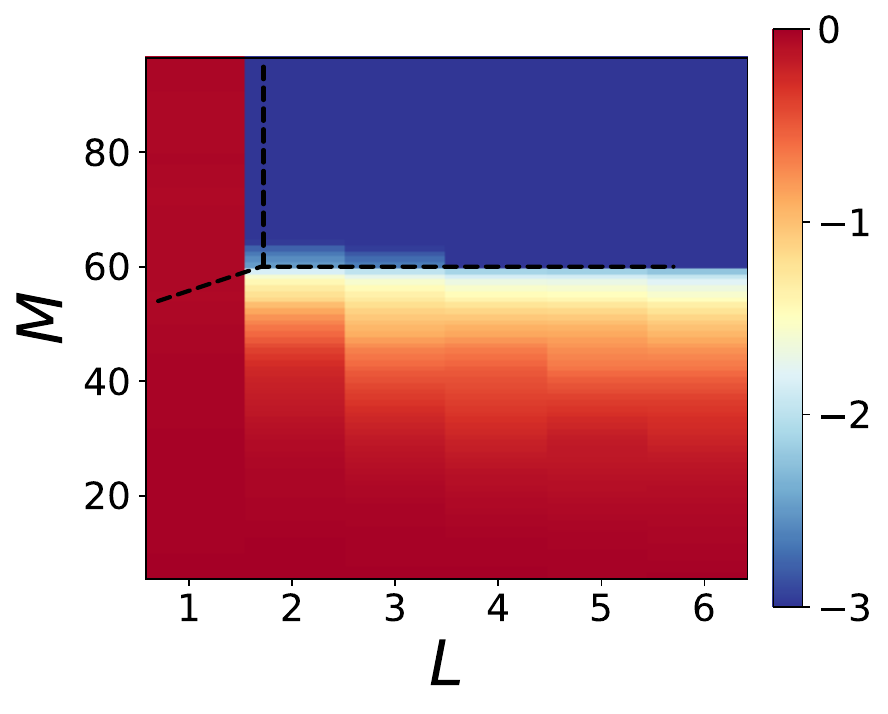}
\subfigimg[width=0.24\textwidth]{c}{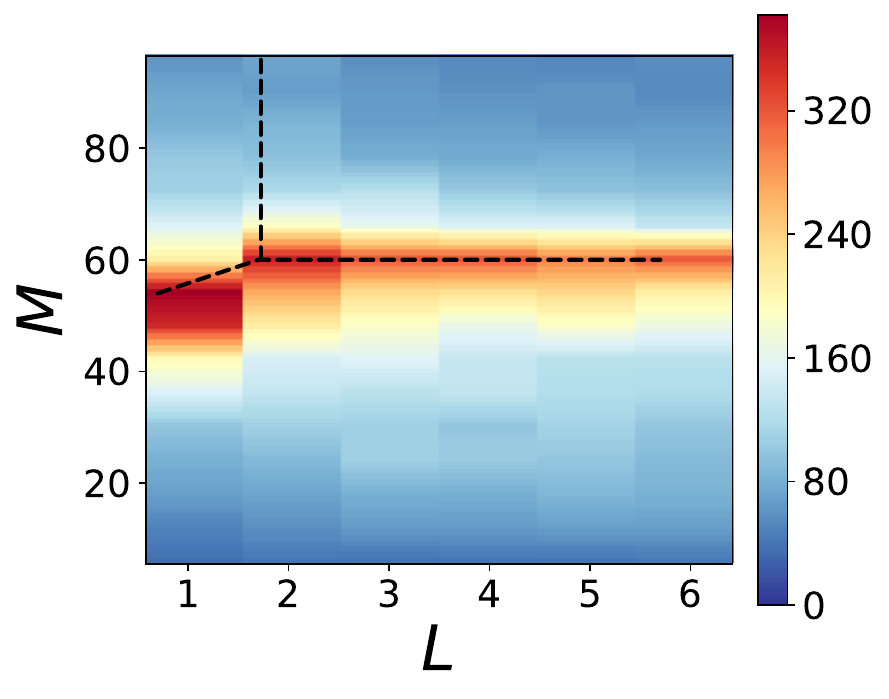}
	\caption{Median error for training $U_\text{Y-CZ}$ with product state $W_{\text{prod}}$ as training data and $N=6$ qubits averaged over 10 instances.
 \idg{a} $C_\text{train}$ against $M$ and $L$.  Color shows logarithm $\log_{10}(C_\text{train})$.
 \idg{b} $C_\text{test}$ against $M$ and $L$.
 \idg{c} Number of training steps $E$ until reaching $C_\text{train}< 10^{-3}$.
	}
	\label{fig:cphase2d}
\end{figure*}

We show numerical results on training with the $U_\text{Y-CZ}$ ansatz (see Fig.~\ref{fig:ansatze_sup}c for definition) in Fig.~\ref{fig:cphase2d}. This model requires $L=2$ states to generalize as we have $R_L\sim N^2$. We find training and test error matches closely the transitions derived from $R_L$ shown as black dashed lines.

\section{DQFIM for overparameterization and generalization for quantum control}\label{sec:control}
The idea of quantum control is to evolve a quantum system by changing the Hamiltonian parameter in time such that it reaches a desired target state.
In particular, we have a time-dependent Hamiltonian~\cite{werschnik2007quantum}
\begin{equation}
    H(x(t))= H_\text{fix} + \sum_{k=1}^K x_k(t) H_k
\end{equation}
with fixed term $H_\text{fix}$ and $K$ time-dependent driving terms $H_k$ with strength $x_k(t)$. 

Commonly, one discretizes the control terms $x_k(t)$ in time. For example, this is the standard approach for gradient ascent pulse engineering algorithm (GRAPE)~\cite{khaneja2005optimal}.
In particular, we discretize time $t$ into $B$ time-steps $\delta t$ where for the $b$th timestep $x_k(b\delta t)$ to $x_k((b+1)\delta t)$ the control parameter is kept constant. This gives us  $M=KB$ driving parameters $\boldsymbol{x}=\{x_{k}(b\delta t)\}_{k=1,b=1}^{K,B}$ with time evolution over total time $T=\delta t B$
\begin{equation}
    U(\boldsymbol{x})=\prod_{b=1}^B\exp(-i H(x(\delta t b)) \delta t)\,.
\end{equation}
Now, the goal is to find the driving terms $x$ such that they realize target states $\{\ket{\psi_\ell}\}_{\ell=1}^L$ starting from initial state $\{\ket{\phi_\ell}\}_{\ell=1}^L$
where we demand $\ket{\psi_\ell}=U(\boldsymbol{x})\ket{\phi_\ell}$ for all $\ell=1,\dots,L$. Common choices for $\ket{\phi_\ell}$ are the basis states $\ket{\ell}$. 

The best control parameters $\boldsymbol{x}$ are found by optimizing them in respect to the fidelity
\begin{equation}
    C_\text{train}(\boldsymbol{x})=1-\frac{1}{L}\sum_{\ell=1}^L \vert\bra{\psi_\ell}U(\boldsymbol{x})\ket{\phi_\ell}\vert^2=1-\frac{1}{L}\sum_{\ell=1}^L \bra{\phi_\ell}U(\boldsymbol{x})^\dagger O_\ell U(\boldsymbol{x})\ket{\phi_\ell}\,.
\end{equation}
where we define the label operator $O_\ell=\ket{\psi_\ell}\bra{\psi_\ell}$ which describes the target state. Thus, the quantum control problem can be mapped onto the unitary learning problem.
Similarly, we can also define a test error when the initial states are drawn from some distribution $W$.

Assuming that the evolution is controllable, i.e. there is a set $\boldsymbol{x}$ of control parameters that can reach the target state, can one find this set of parameters via an optimization program minimizing $C_\text{train}(\boldsymbol{x})$?  
It is known that overparameterized problems where one has a  large number of controls $K$ and many timesteps $B$ are easily able to find a good solution, while training of underparameterized control models tends to get stuck in local minimas and it is hard to find good solutions~\cite{rabitz2004quantum,bukov2018reinforcement}. However, how to choose $B=M/K$ to achieve overparameterization for a given model and initial state? Obviously, if $B$ is chosen to small, then it is hard to find a good control protocol, while choosing $B$ too large will yield  a control protocol that is more complex than necessary. 
The rank of the DQFIM tell us at what $B$ the model overparameterizes, which is the best choice of discretization.

We now demonstrate two control problems and show that the maximal rank of the DQFIM allows us to determine the overparameterization transition as function of $M$ and $L$.

\begin{figure*}[htbp]
	\centering	
\subfigimg[width=0.24\textwidth]{a}{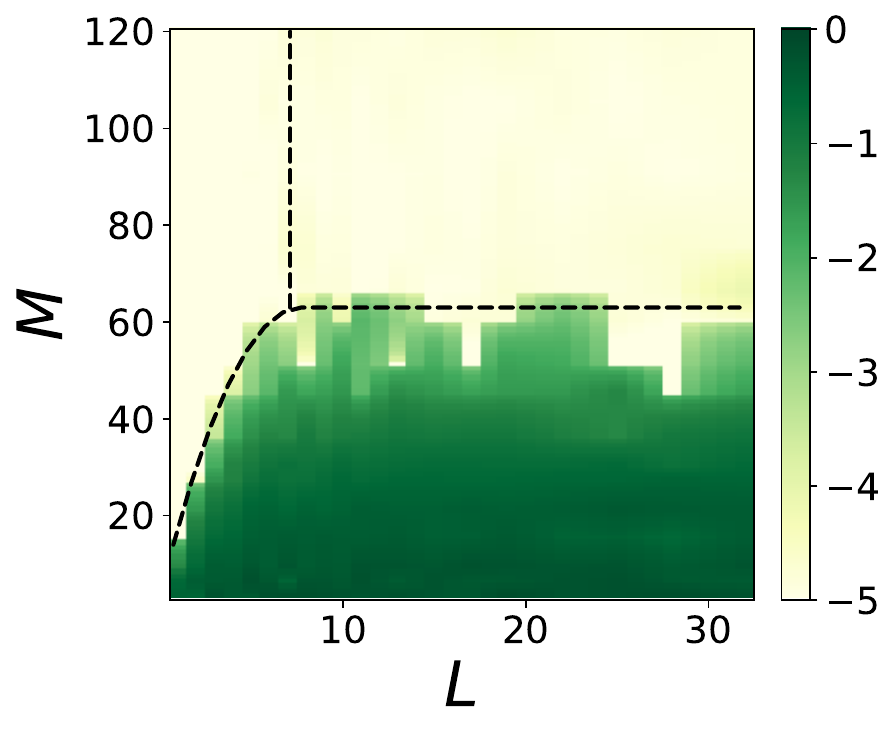}
\subfigimg[width=0.24\textwidth]{b}{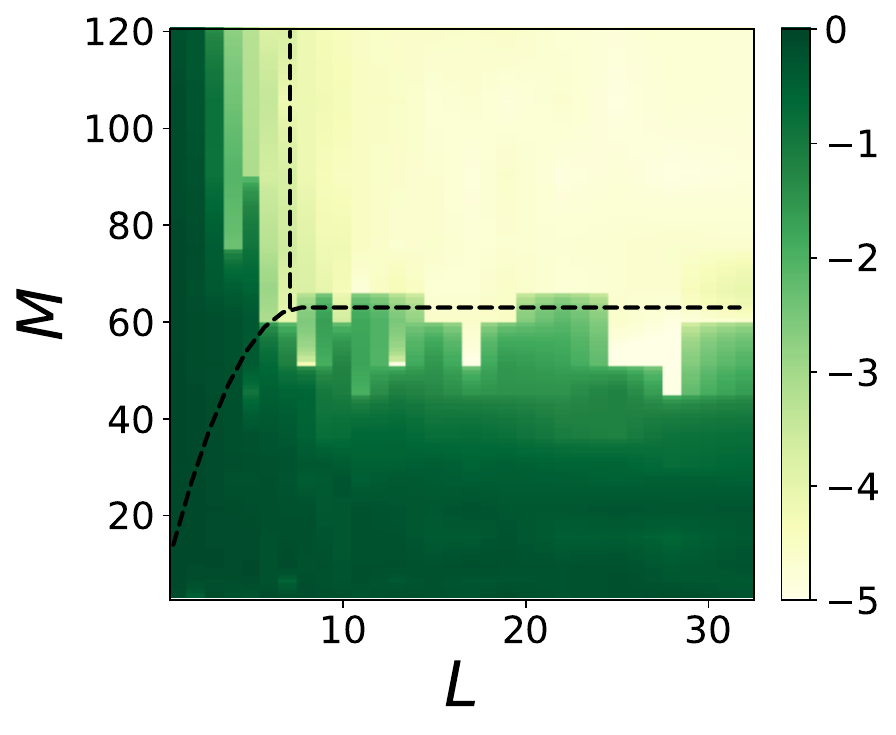}
\subfigimg[width=0.24\textwidth]{c}{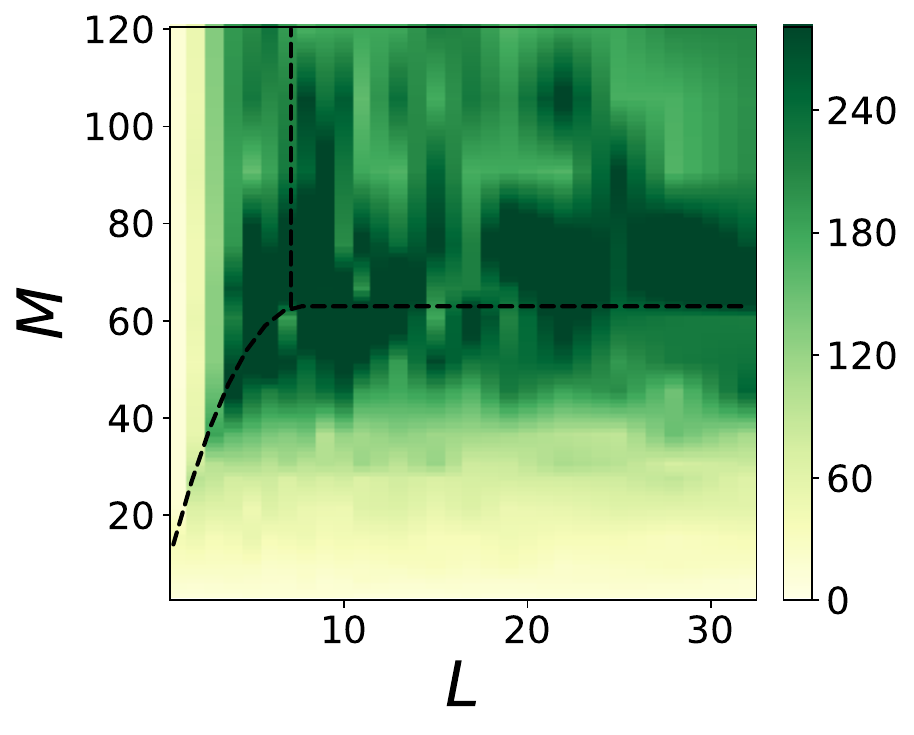}
	\caption{Fidelity of optimized control protocol $\boldsymbol{x}$ for non-integrable Ising Hamiltonian~\eqref{eq:controlIsing}. 
 \idg{a} Median of trained cost function $\log(C_\text{train})$ against number of control parameters $M$ and number of training states $L$. Black dashed line is maximal rank of DQFIM $R_L$.
 \idg{b} Median of test cost $\log(C_\text{test})$.
 \idg{c} Number of training steps $E$ until reaching $C_\text{train}<10^{-4}$.
We have $N=3$ qubits, $g=0.3$, total time $T=6$ and initial states are random product states and target states are chosen randomly from the feasible control space.
	}
	\label{fig:controlIsing}
\end{figure*}

First, we study an Ising Hamiltonian with additional longitudinal field that is known to be non-integrable with the choice 
\begin{equation}\label{eq:controlIsing}
    H_\text{fix}=\sum_{n=1}^N \sigma_n^x \sigma_{n+1}^x + g\sigma_n^x
\end{equation}
and the tunable transversal field $H_k=x_k(t)\sigma_k^z$ with $k=1,\dots, N=K$. We show the result in Fig.~\ref{fig:controlIsing}.
We show the training error in Fig.~\ref{fig:controlIsing}a. We find that the training error vanishes beyond a certain $M=B/K$, which his given by the maximal rank of the DQFIM which is shown as dashed line. This is the transition to overparamterization, which also depends on the number of initial states $L$ for the control problem.
in Fig.~\ref{fig:controlIsing}b we show the test error against arbitrarily chosen initial states. The vertical dashed line shows the critical number of initial states $L=2R_\infty/R_1$, which we find matches the transition in the text error accurately.
In Fig.~\ref{fig:controlIsing}c we show the number of iterations until the training converges. We find that around $M\sim M_\text{c}$, the number of iterations needed becomes much larger.

We also study the Heisenberg XXZ Hamiltonian, which is non-integrable but has symmetries which restrict the dynamics to a subspace of the full Hilbertspace~\cite{gibbs2022dynamical}.
It is given by
\begin{equation}\label{eq:controlHeisenberg}
    H_\text{fix}=\sum_{n=1}^N \sigma_n^x \sigma_{n+1}^x +\sigma_n^y \sigma_{n+1}^y + g\sigma_n^z \sigma_{n+1}^z
\end{equation}
and tunable field $H_k=x_k(t)\sigma_k^z$. 

\begin{figure*}[htbp]
	\centering	
\subfigimg[width=0.24\textwidth]{a}{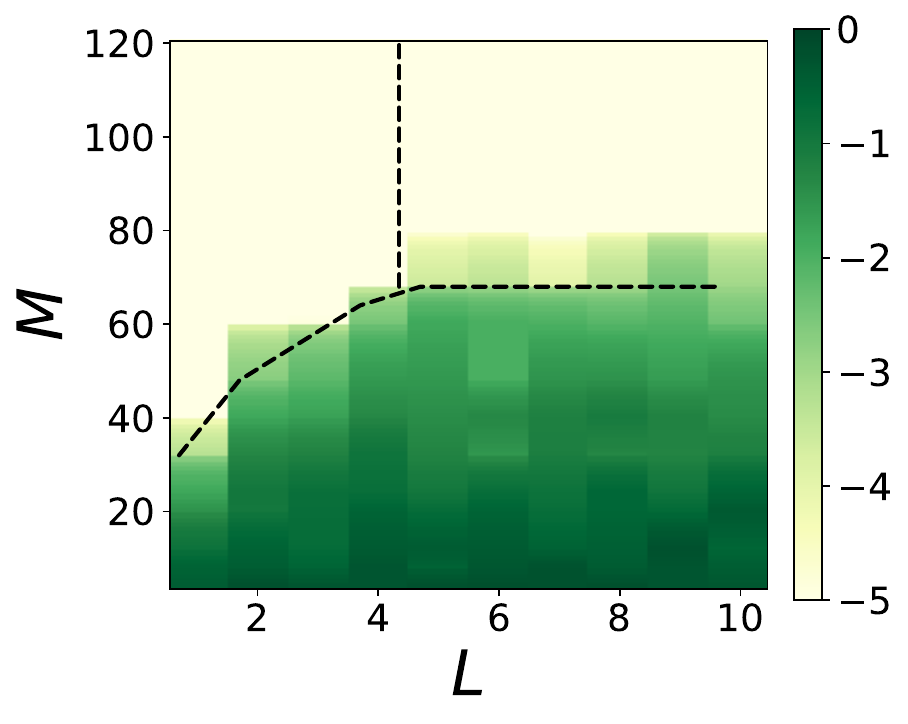}
\subfigimg[width=0.24\textwidth]{b}{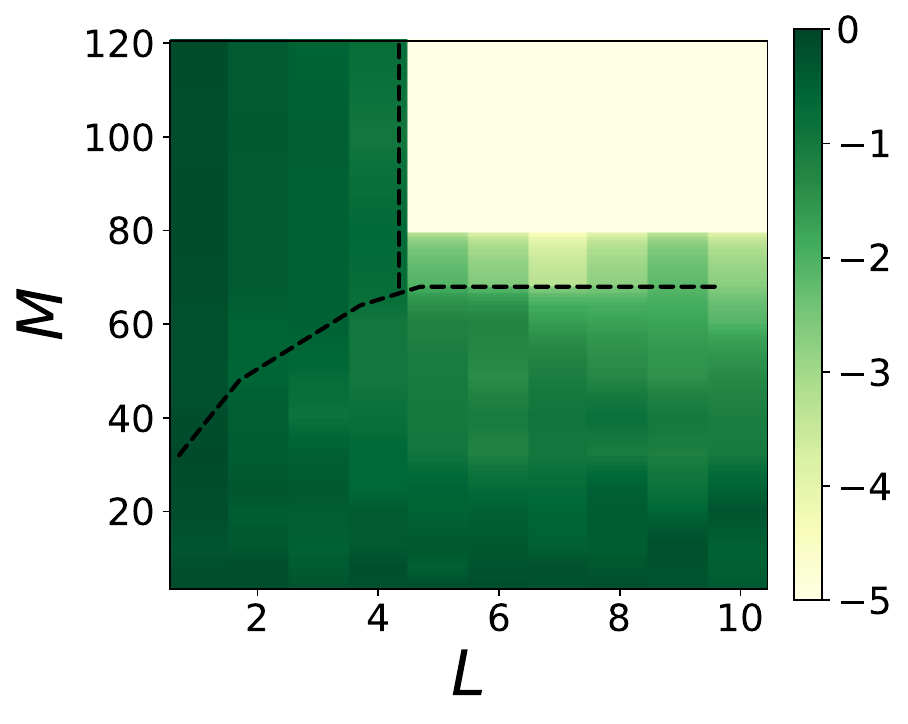}
\subfigimg[width=0.24\textwidth]{c}{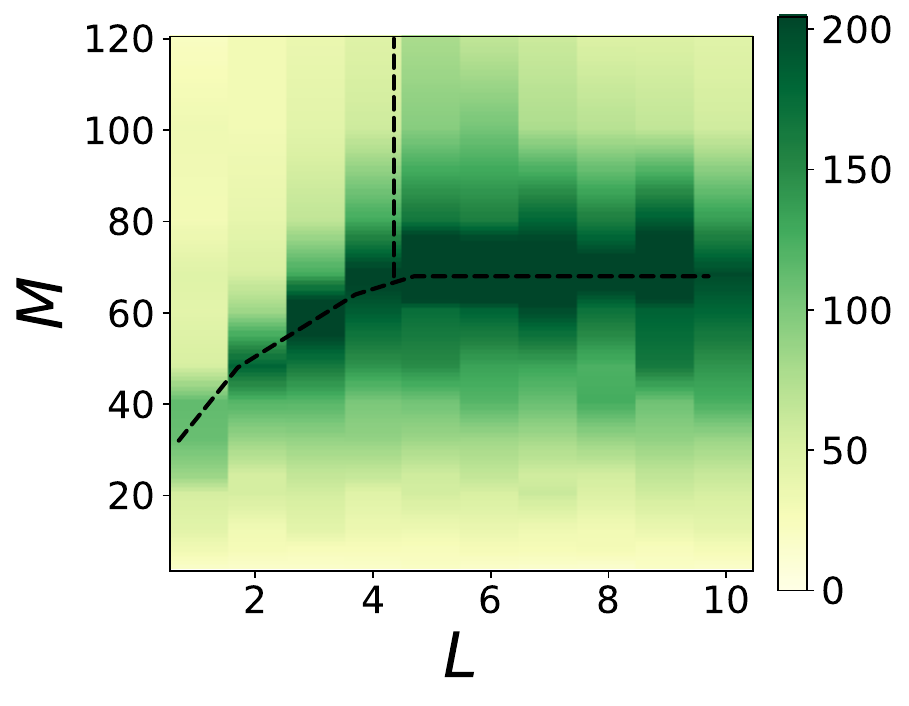}
	\caption{Fidelity of optimized control protocol $\boldsymbol{x}$ for Heisenberg Hamiltonian~\eqref{eq:controlHeisenberg}. 
 \idg{a} Median of trained cost function $\log(C_\text{train})$ against control parameters $M$ and number of training states $L$. Black dashed line is $M_\text{c}$ from maximizing rank of DQFIM $R_L$.
 \idg{b} Median of test cost $\log(C_\text{test})$.
 \idg{c} Number of training steps $E$ until reaching $C_\text{train}<10^{-4}$.
We have $N=4$ qubits, $g=0.3$, total time $T=12$ and initial states are random product states and target states are chosen randomly from the feasible control space.
	}
	\label{fig:controlHeisenberg}
\end{figure*}

We show the result for the Heisenberg Hamiltonian in Fig.~\ref{fig:controlHeisenberg}. 
We show the training error after training against $L$ and number of control parameters $M$ in Fig.~\ref{fig:controlHeisenberg}a. The black dashed line is the maximal rank of the DQFIM, which matches when the training error vanished due to overparameterization. Note that while the Heisenberg model is non-integrable, due its symmetries we require less parameters $M$ to overparameterize than for a generic Hamiltonian (e.g. for $N=4$ one has overparameterization for $M\ge R_5=134$ for $L=5$, while for Heisenberg model we find overparameterization at much lower $M$).
We show the test error in Fig.~\ref{fig:controlHeisenberg}b, where we find that the test error vanishes for $L\ge5$,, which matches the vertical dashed line which shows the critical $L_\text{crit}=2R_\infty/R_1$. 
In Fig.~\ref{fig:controlHeisenberg}c, we show the number of iterations until the training converges. We find that close to critical $M_\text{c}$ and $L_\text{c}$ more iterations are needed to successfully train.

In conclusion, we find that the rank of the DQFIM precisely tells how to discretize control parameters to overparameterize quantum control problems.

\section{Overparameterization for generative models}\label{sec:generative}
In generative modeling, the goal is to produce samples $x\sim p(x)$ from a probability distribution $p(x)$. 
In quantum generative learning, called quantum circuit born machines, $x$ is sampled from a quantum state $\ket{\psi(\boldsymbol{\theta})}$ with $p(x)\approx \vert \braket{x}{\psi(\boldsymbol{\theta})} \vert^2$~\cite{coyle2020born}. The state $\ket{\psi(\boldsymbol{\theta})}$ is trained from samples of the distribution $\{x_\ell\}_{\ell=1}^L$. 
Various methods to write down cost functions are known which are reviewed in Ref.~\cite{rudolph2023trainability}. 

Here, we highlight what is called fidelity training, an explicit learning method which can evade barren plateaus for shallow circuits~\cite{rudolph2023trainability}.
In this method, the cost function is written as
\begin{equation}
    C_\text{train}(\boldsymbol{\theta})=\bra{\psi}U^\dagger(\boldsymbol{\theta}) \ket{\Phi}\bra{\Phi}U(\boldsymbol{\theta})\ket{\psi}
\end{equation}
with state $\ket{\Phi}=\sum_{\ell=1}^L \sqrt{q(x_\ell)}\ket{x_\ell}$ encoding the training set where $\ket{x_\ell}$ is the basis state corresponding to $x_\ell$ and $q(x_\ell)$ is the multiplicity of each outcome $x_\ell$. 
Note that this scheme corresponds to learning a unitary with a single $L=1$ training state and label operator $O=\ket{\Phi}\bra{\Phi}$, matching the canonical quantum compiling task.
Thus, we can immediately give the overparameterization threshold for fidelity training of quantum generative models as $M_\text{c}\sim R_1$. Note as the full classical dataset is encoded into a single state, there is no explicit $L$ dependence for overparameterization of fidelity training of quantum generative models.

\section{Overparameterization of variational quantum eigensolver for eigenstates}\label{sec:excited}
The variational quantum eigensolver (VQE) finds the ground state of a Hamiltonian $H$~\cite{bharti2021noisy}. This algorithm has been extended to find $L$ low-energy eigenstates of $H$ via the subspace-search variational quantum eigensolver (SSVQE)~\cite{nakanishi2019subspace}. 
The idea of SSVQE is to select $M$-parameter ansatz $U(\boldsymbol{\theta})$,  prepare $L$ orthogonal basis states $\{\ket{k}\}_{k=1}^L$ and minimize the cost function
\begin{equation}\label{eq:SSVQE}
    C(\boldsymbol{\theta})=\sum_{k=1}^L w_k \bra{k}U(\boldsymbol{\theta})^\dagger HU(\boldsymbol{\theta})\ket{k}
\end{equation}
where the weights $w_k\in(0,1)$ are chosen as $w_{k} > w_{k+1}$. One can easily see that $C$ is minimal when for all $k=1,\dots,L$ we have $\ket{\psi_k}=U(\boldsymbol{\theta})\ket{k}$, where $\ket{\psi_k}$ is the eigenstate corresponding to the $k$-th lowest eigenvalue of $H$. Thus, by minimizing $C(\boldsymbol{\theta})$ we can find all $L$ lowest eigenstates of $H$.

We now study overparameterization of the SSVQE as function of the number $L$ of low-energy eigenstates to be found.
For the case of finding only the ground state, i.e. $L=1$, it is known that a generic VQE circuit ansatz becomes overparameterized for $M\sim 2^N$ parameters where training converges with high probability to the ground state~\cite{anschuetz2021critical}.

We now use the DQFIM to determine overparameterization for $L>1$ which to our knowledge has not been known. We assume that the Hamiltonian is non-degenerate so we can properly order eigenvalues. Then, we observe that training the SSVQE corresponds to a unitary learning problem where we must find the $U(\boldsymbol{\theta})$ that maps the $k=1,\dots,L$ initial states $\ket{k}$ to their unique target states $\ket{\psi_k}=U(\boldsymbol{\theta})\ket{k}$. 
Thus, we argue that training the SSVQE requires the same critical number of parameters $M_\text{c}(L)$ for overparameterization for unitary learning problem. 
In particular, we have 
\begin{equation}
    M_\text{c}(L)=R_L(\rho_L,U(\boldsymbol{\theta},M))
\end{equation}
where $R_L$ is the maximal rank of the DQFIM and $\rho_L=L^{-1}\sum_{k=1}^L \ket{k}\bra{k}$.
For a generic ansatz unitary (for example composed of layers of parameterized single qubit rotations and CNOT gates), we get as derived in Sec.~\ref{sec:RLexact}
\begin{equation}
    M_\text{c}^\text{generic}(L)=2dL-L^2-1\,.
\end{equation}
where $d=2^N$ is the Hilbertspace dimension.

\begin{figure*}[htbp]
	\centering	
\subfigimg[width=0.24\textwidth]{a}{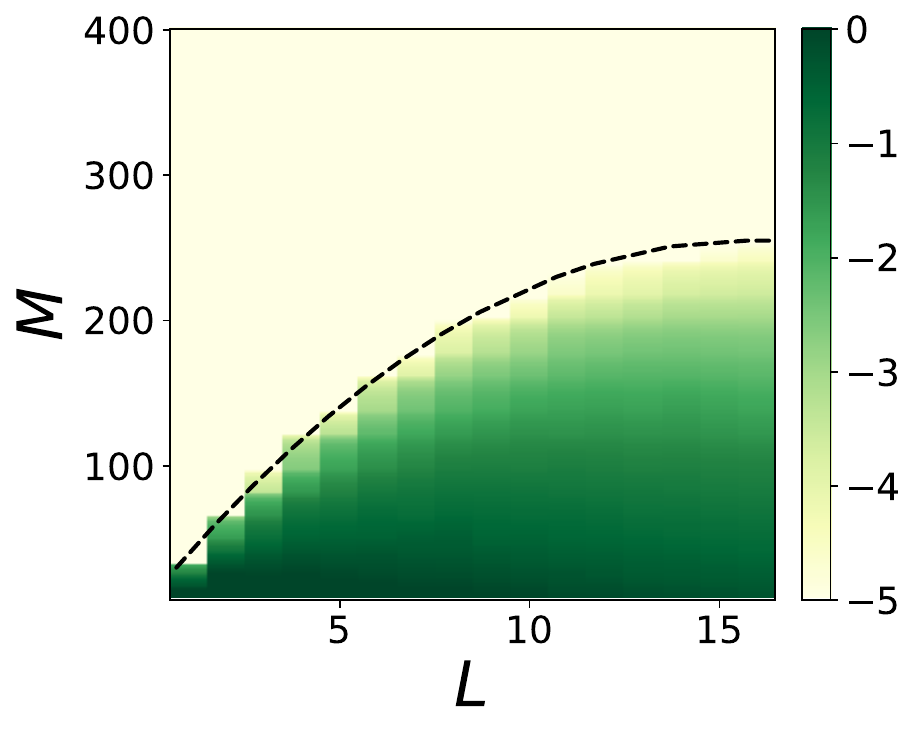}
\subfigimg[width=0.24\textwidth]{b}{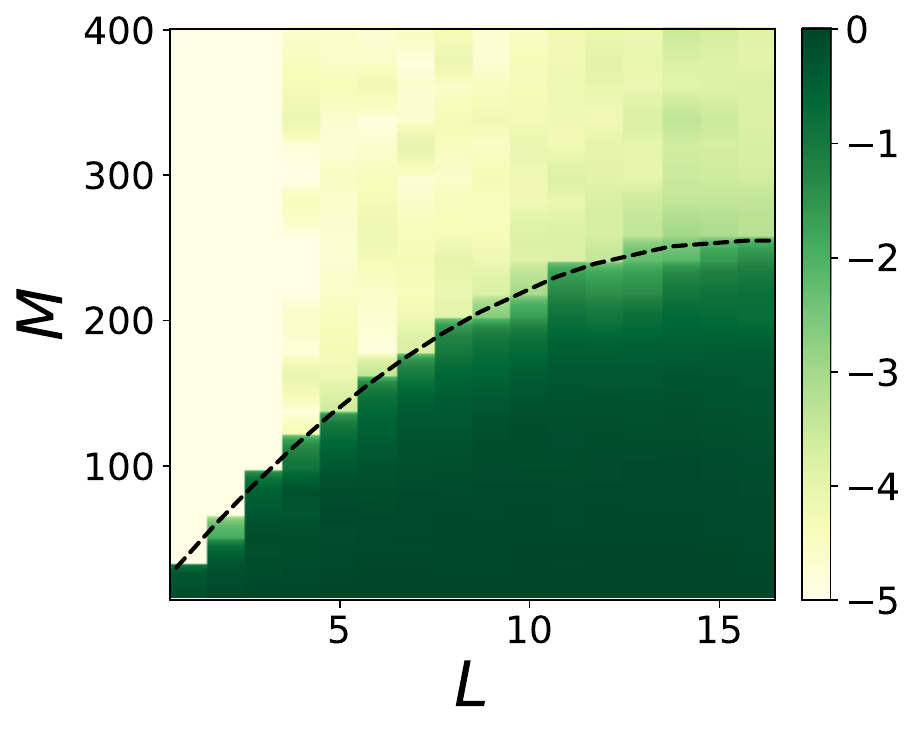}
\subfigimg[width=0.24\textwidth]{c}{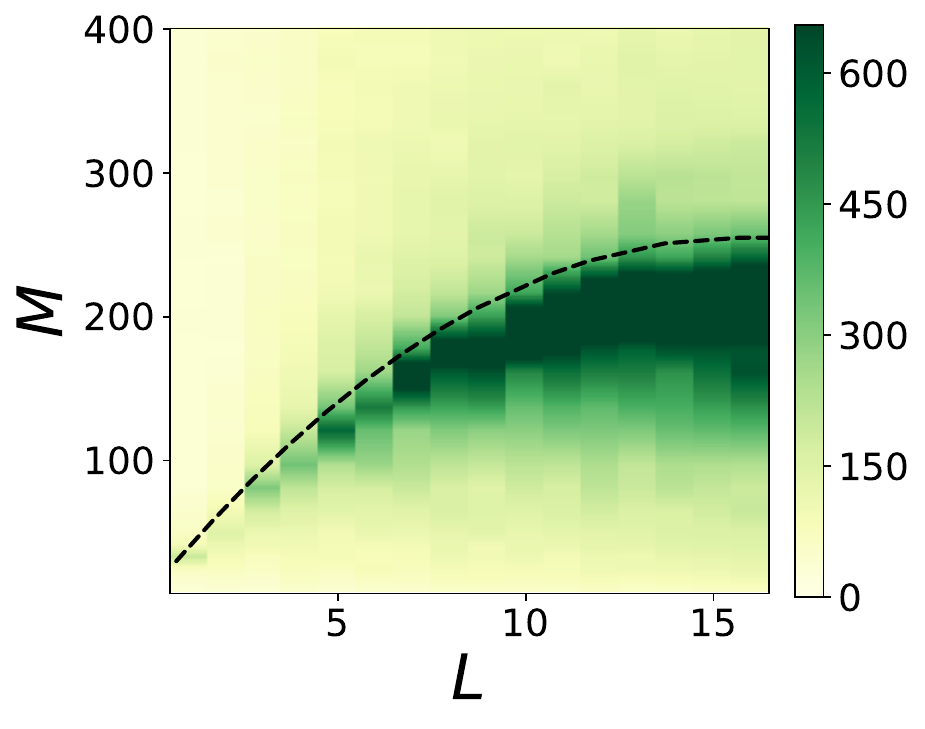}
	\caption{Performance of SSVQE for finding lowest eigenstates of Hamiltonian~\eqref{eq:disorder}.
 \idg{a} Median of trained cost function $\log(C_\text{train}-C_\text{min})$~\eqref{eq:SSVQE} against circuit parameters $M$ and number of eigenstates $L$. Dashed line is maximal rank of DQFIM $R_L$.
 \idg{b} Median of fidelity $\log(F)$.
 \idg{c} Number of training steps $E$ until reaching $C_\text{train}<10^{-4}$.
    We use generic ansatz $U_\text{HE}$ of Fig.\ref{fig:ansatze}a with $N=4$ qubits.
	}
	\label{fig:SSVQE}
\end{figure*}

We now show numerical results on the SSVQE. We learn the $L$ eigenstates with lowest energies of the Hamiltonian
\begin{equation}\label{eq:disorder}
    H=\sum_{j} \sigma^x_j\sigma^x_{j+1}+ g\sigma^x_{j}+ h_j \sigma^z_{j}
\end{equation}
with $g=0.2$ and randomly chosen $h_j\in\{-1,1\}$.
We minimize the cost function~\eqref{eq:SSVQE} until convergence (to either global or local minima).
To evaluate the performance of the SSVQE, we compute the fidelity of the found states with the exact eigenstates via
\begin{equation}
    F=\frac{1}{L}\sum_{k=1}^L \vert \bra{\psi_k}U(\boldsymbol{\theta}^\ast)\ket{k}\vert^2
\end{equation}
where $\boldsymbol{\theta}^\ast$ is the parameter set found via training, $\ket{\psi_k}$ the eigenstate of $H$ corresponding to the $k$th lowest eigenvalue. 
We show the converged function (minus the theoretical minimal possible value) as function of $M$ and $L$ in Fig.~\ref{fig:SSVQE}a. The dashed line is the maximal rank of the DQFIM $R_L$. We find that the cost function becomes very small once $M>R_L$, indicating the overparameterization transition. We show $F$ in Fig.~\ref{fig:SSVQE}b, observing that the fidelity converges to $1$ for $M>R_L$.
Finally, we plot the number of training iterations needed to converge in Fig.~\ref{fig:SSVQE}c. We find that close to the overparameterization transition the training time increases, indicating that at the transition the optimization landscape is more complex~\cite{bukov2018reinforcement}.

\section{Overparameterization and generalization in classification tasks}\label{sec:classify}
We now consider a classification task with quantum machine learning. This is a supervised learning algorithm where one is given a classical training dataset $\{y_\ell,x_\ell \}$ with feature vector $x_\ell$ and binary label $y_\ell\in\{-1,+1\}$. 
The goal is to use the quantum computer to classify the data. 
This can be achieved by encoding the feature vector $x_\ell$ into a quantum state $\ket{\psi(x_\ell)}\equiv\ket{\psi_\ell}$. Then, one learns a parameterized unitary $U(\boldsymbol{\theta})$ such that the expectation value $\bra{\psi_\ell} O_\ell\ket{\psi_\ell}$ in respect to an observable $O_\ell$ reproduces the label, i.e. we want to find $\boldsymbol{\theta}$ such that $\bra{\psi_\ell}U^\dagger(\boldsymbol{\theta}) O_\ell U(\boldsymbol{\theta})\ket{\psi_\ell}=y_\ell$ for all $\ell=1,\dots,L$. 
A popular choice is $O_\ell=y_\ell\sigma^z$ with $z$ Pauli operator $\sigma^z$. The corresponding quantum dataset is $S_L=\{\ket{\psi_\ell},y_\ell \sigma^z\}_{\ell=1}^L$.
The quantum classifier can be trained by minimizing the cost function~\cite{farhi2018classification}
\begin{equation}
    C_\text{train}(\boldsymbol{\theta})=1-\frac{1}{L}\sum_{\ell=1}^L \bra{\psi_\ell}U^\dagger(\boldsymbol{\theta}) O_\ell U(\boldsymbol{\theta})\ket{\psi_\ell}\,.
\end{equation}
After training, the classifier is tested against unseen test data drawn from some distribution $W$ to determine its generalization capability.
For this, we can use the test error
\begin{equation}\label{eq:test_classify}
    C_\text{test}(\boldsymbol{\theta})=1-\frac{1}{L}\sum_{\{\ket{\psi}\}\in W}^L \bra{\psi}U^\dagger(\boldsymbol{\theta}) O_{\psi} U(\boldsymbol{\theta})\ket{\psi}\,
\end{equation}
with $O_{\psi}=y_{\psi}\sigma^z$ where $y_{\psi}$ is the corresponding label to $\ket{\psi}$.
When $C_\text{test}\approx 0$, we can optimally classify any test data, i.e. we are able to predict the correct label with only a single sampling of $\bra{\psi}U^\dagger(\boldsymbol{\theta}) O_{\psi} U(\boldsymbol{\theta})\ket{\psi}$.

What is a good choice for $O$ to classify two classes of data?
For the unitary learning problem $O=V\ket{\psi}\bra{\psi}V^\dagger$ is a projector onto the target to be learned. Thus there is only one correct target state with $V\ket{\psi}=U(\boldsymbol{\theta})\ket{\psi}$ which is not useful for classification tasks. 
Commonly, for binary classification tasks one chooses $O$ to be a Pauli operator, e.g. $O=\sigma^z_1$~\cite{farhi2018classification}. This operator has two eigenvalues $+1$,$-1$ which can be used to distinguish the two classes. Now, one trains $U(\boldsymbol{\theta})$ such that it transforms data with $y=1$ label to the $+1$ eigenspace of $O$, while data with $y=-1$ is transformed into the $-1$ eigenspace. 
Now, the solution space is highly degenerate as there are $2^{N-1}$ degenerate eigenvalues for each label.
This makes the classification problem easier to learn in contrast to the unitary learning problem (which has only one correct solution), and classification problems require less parameters to overparameterize. 
Thus, the critical number of parameters $M_\text{c}(L)$ from the DQFIM presents an upper bound on overparameterization for classification tasks. 

In the following, we study overparameterization numerically for classification. As we will see, the critical number of parameters $M_c^{\text{classify}}$ is reduced compared to learning unitaries. In particular, we observe numerically that for $\sigma^z_1$ we require $M_\text{c}^{\text{classify}}= R_L \gamma$ to overparameterize, where $\gamma\le1$. 
We observe that $\gamma$ depends on the chosen circuit ansatz. In particular, for a generic ansatz circuit and $\sigma^z$ we find $\gamma=\frac{1}{2}$, while for a particle-number conserving $U_\text{XY}$ circuit we find $\gamma=\frac{1}{3}$.

\begin{figure*}[htbp]
	\centering	
\subfigimg[width=0.24\textwidth]{a}{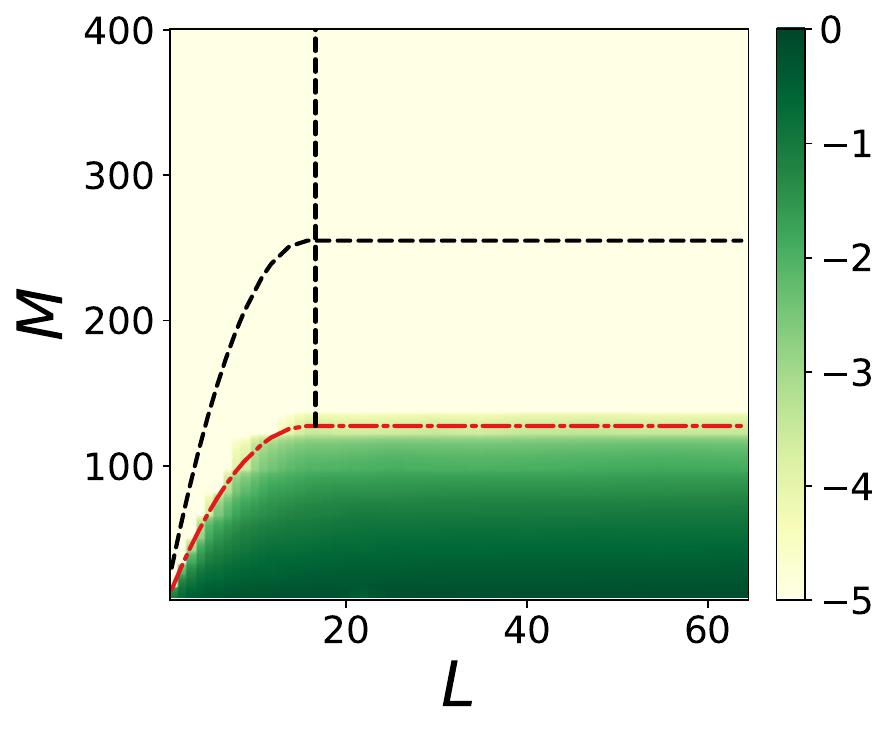}
\subfigimg[width=0.24\textwidth]{b}{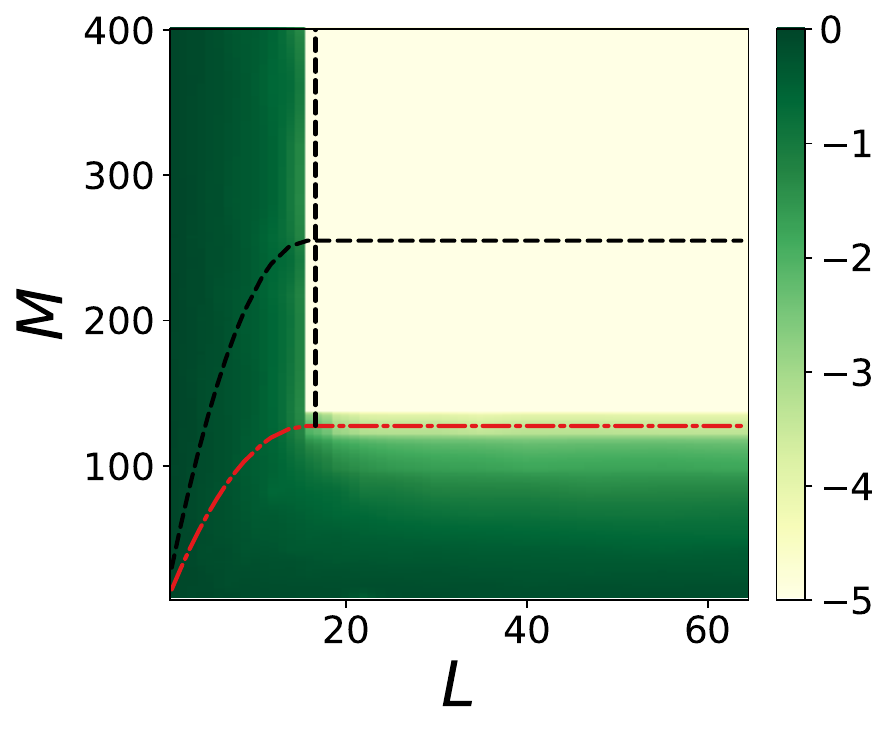}
\subfigimg[width=0.24\textwidth]{c}{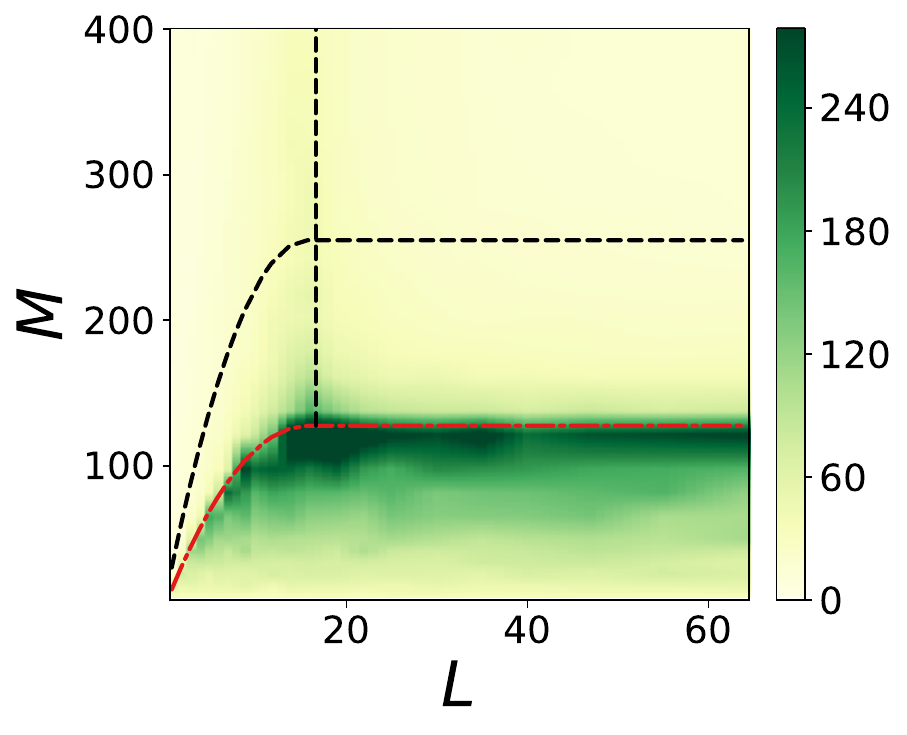}
	\caption{Performance of classification task for generic circuit ansatz.
 \idg{a} Mean of trained cost function $\log(C_\text{train})$ against circuit parameters $M$ and number of training states $L$. Black dashed line is maximal rank of DQFIM $R_L$, while red line is $R_L\gamma$ with empirical factor $\gamma=1/2$.
 \idg{b} Median of test cost $\log(C_\text{test})$.
 \idg{c} Number of training steps $E$ until reaching $C_\text{train}<10^{-4}$.
    We use generic ansatz $U_\text{HE}$ of Fig.\ref{fig:ansatze}a with $N=4$ qubits, $O=\sigma^z_1$ as classification operator. States with label $y=1$ are randomly drawn from a subspace with eigenvalue $+1$ of rotated operator $V\sigma^z_1 V^\dagger$, while $y=-1$ from the $-1$ subspace.
	}
	\label{fig:classifyCost}
\end{figure*}

We study the classification task in Fig.~\ref{fig:classifyCost} for a generic ansatz circuit $U_\text{HE}$. In Fig.\ref{fig:classifyCost}a, we show the classification training error $C_\text{train}$ against circuit parameters $M$ and training states $L$. We find that the training error becomes small when $M\ge R_L/2$ which is plotted as red line. This is half of the usual maximal rank of the DQFIM $R_L$ which is also shown as black dashed line. While $R_L$ describes overparameterization for unitary learning problems, for classification tasks with Pauli operators we find that overparameterization requires only half the number of circuit parameters. This is due to the degeneracy  of the Pauli operator $O=\sigma^z_1$.

In Fig.\ref{fig:classifyCost}b, we show the test error. We find generalization for $L\ge16$, which is the same as for the unitary learning task.
Finally, in Fig.\ref{fig:classifyCost}c we show the number of training iterations needed to train. We find a peak in iterations for $M=R_L/2$.

\begin{figure*}[htbp]
	\centering	
\subfigimg[width=0.24\textwidth]{a}{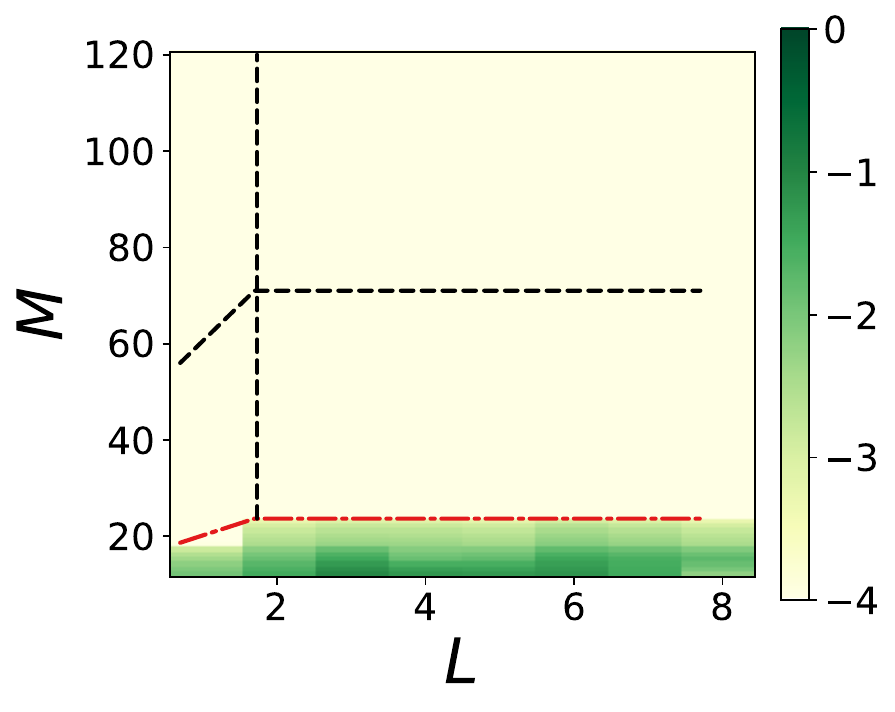}
\subfigimg[width=0.24\textwidth]{b}{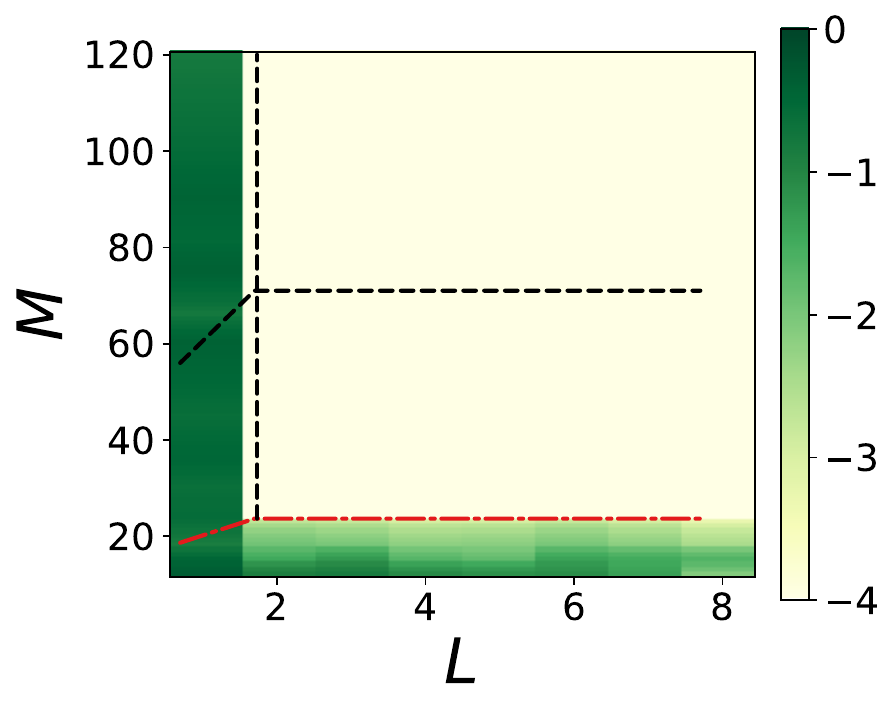}
	\caption{Performance of classification task for particle-number conserving $U_{XY}$ circuit.
 \idg{a} Mean of trained cost function $\log(C_\text{train})$ against circuit parameters $M$ and number of training states $L$. Black dashed line is maximal rank of DQFIM $R_L$, while red line is $R_L\gamma$ with empirical factor $\gamma=1/3$.
 \idg{b} Median of test cost $\log(C_\text{test})$.
 \idg{c} Number of training steps $E$ until reaching $C_\text{train}<10^{-4}$.
    We use $U_{XY}$ ansatz of Fig.\ref{fig:ansatze}b with $N=6$ qubits and $O=\sigma^z_1$ as classification operator. States with label $y=1$ are randomly drawn from a subspace with eigenvalue $+1$ of rotated operator $V\sigma^z_1 V^\dagger$, while $y=-1$ from the $-1$ subspace.
	}
	\label{fig:classifyCostIswap}
\end{figure*}

Next, in Fig.~\ref{fig:classifyCostIswap} we study classification using the particle-number conserving $U_\text{XY}$ ansatz of Fig.~\ref{fig:ansatze}b. For this ansatz overparameterization scales polynomially with qubit number. In Fig.~\ref{fig:classifyCostIswap}a we show the training cost against $M$, $L$. We find that training cost becomes small for $M\ge R_L/3$, which indicates a factor $\gamma=1/3$ reduction compared to unitary learning. For the test error Fig.~\ref{fig:classifyCostIswap}b we find generalization for $L\ge2$, which is the same as for unitary learning. 

Note that here we evaluated the test error by looking at $C_\text{test}$. 
One can also regard test error as the misclassifaction rate for the following decision rule: When $\bra{\psi}U^\dagger(\boldsymbol{\theta}) O_{\psi} U(\boldsymbol{\theta})\ket{\psi}\geq\epsilon$ we classify $y=1$, and for $\bra{\psi}U^\dagger(\boldsymbol{\theta}) O_{\psi} U(\boldsymbol{\theta})\ket{\psi}\leq-\epsilon$ we classify as $y=-1$, with some threshold $\epsilon$. 
$C_\text{test}$ as defined in~\eqref{eq:test_classify} corresponds to the rule $\epsilon=1$. Smaller $\epsilon$ can yield smaller generalization errors, however at the cost of increasing the number of measurement shots on the quantum computer $N_\text{sample}\sim 1/\epsilon^2$ as one needs to evaluate the observable $\bra{\psi}U^\dagger(\boldsymbol{\theta}) O_{\psi} U(\boldsymbol{\theta})\ket{\psi}$ to higher precision. 
The DQFIM gives us the generalization threshold $L_\text{c}\sim 2R_\infty/R_1$ for the case $\epsilon=1$, \revB{i.e. one uses only a single measurement shot to classify per datapoint~\cite{recioarmengol2024singleshot}.} For smaller $\epsilon$, $L_\text{c}(\epsilon)\leq L_text{c}(\epsilon=1)$ is expected to decrease. Thus, generalization bound via the DQFIM can be seen as an upper bound on the number of datapoints needed to generalize.

\section{Generalization and empirical generalization error}\label{sec:risk}
We define generalization via the error of the cost function $C_\text{test}$ averages over the full data ensemble $W$.  In our studies, we choose the problem such that $C_\text{test}=0$ can be achieved for at least one parameter $\boldsymbol{\theta}_\text{g}$.

\begin{figure*}[htbp]
	\centering	
\subfigimg[width=0.28\textwidth]{a}{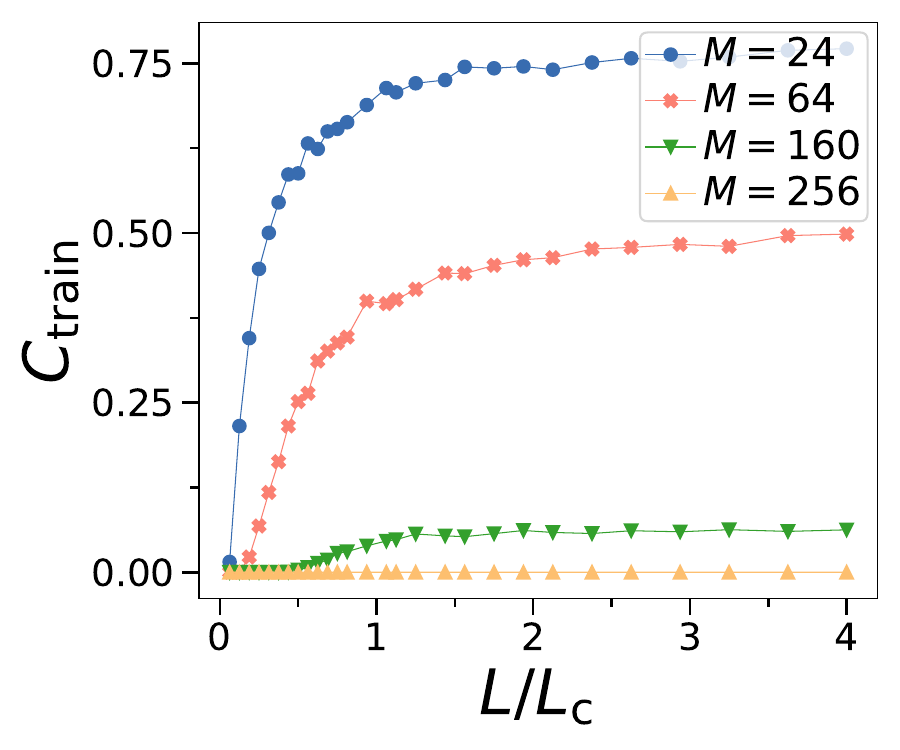}
\subfigimg[width=0.28\textwidth]{b}{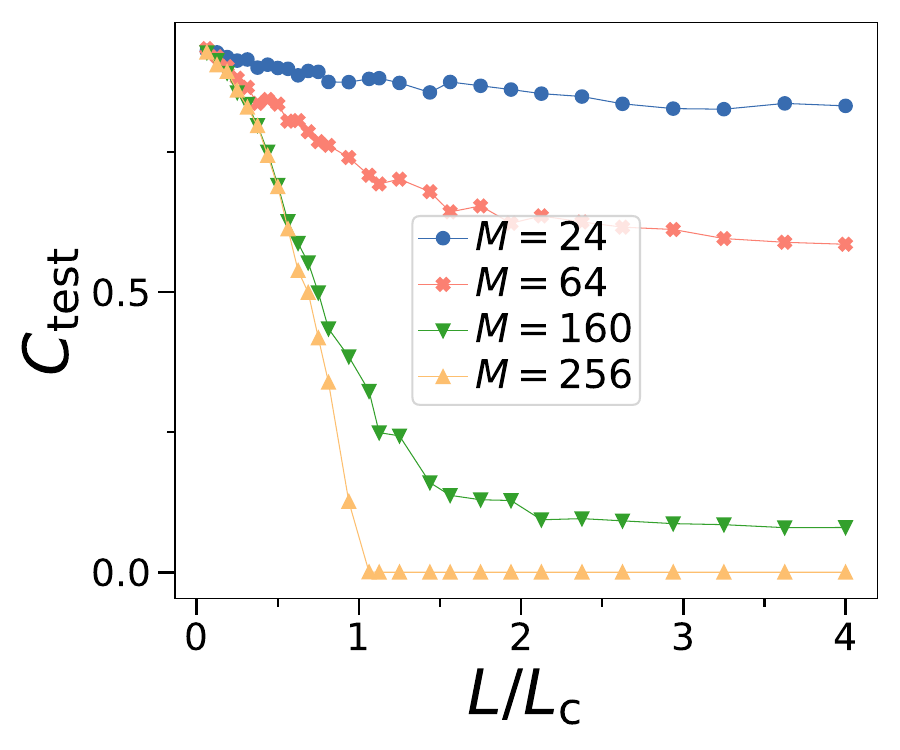}
\subfigimg[width=0.28\textwidth]{c}{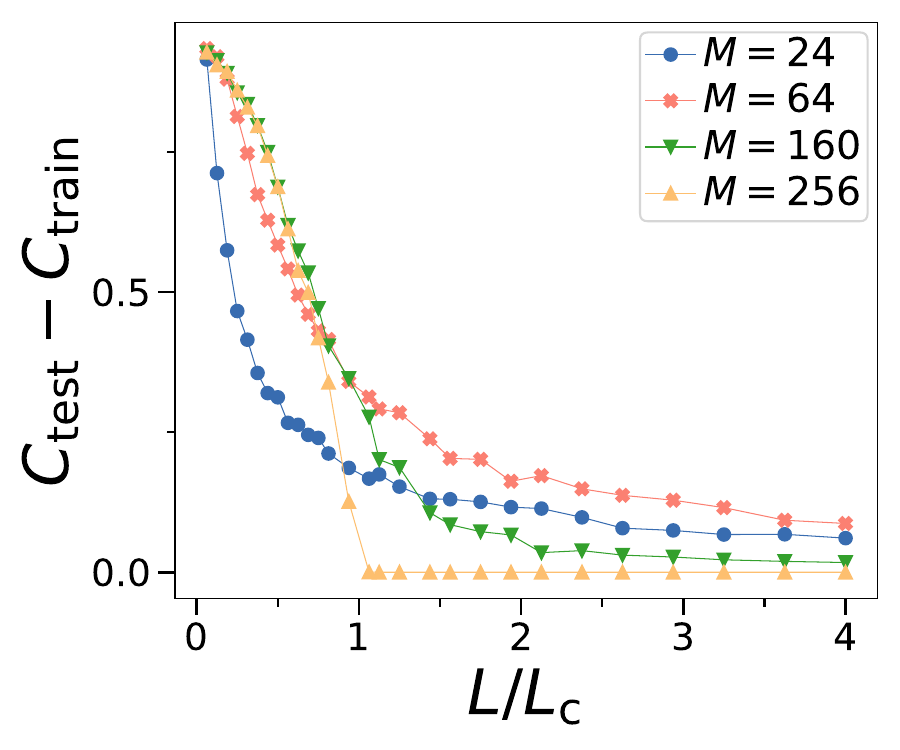}
	\caption{Training error, test error and empirical generalization error for hardware-efficient ansatz with haar random training states for $N=4$ qubits. We assume that there is an optimal solution, i.e. there is at least one parameter with $C_\text{test}(\boldsymbol{\theta_\text{g}})=0$.
 \idg{a} $C_\text{train}$ against  $L$ for different $M$.
 \idg{b} $C_\text{test}$ against  $L$ for different $M$.
 \idg{c} Empirical generalization error $C_\text{test}$-$C_\text{test}$ against  $L$ for different $M$.
	}
	\label{fig:risk}
\end{figure*}

In general, the minimal achievable $C_\text{test}$ for given ansatz and data distribution may not be known. 
Thus, often the empirical generalization error $\zeta=C_\text{test}-C_\text{train}$ of the trained model is used as a proxy to evaluate generalization~\cite{caro2021generalization}. 
In Fig.~\ref{fig:risk}, we compare $C_\text{train}$, $C_\text{test}$ and empirical generalization error $C_\text{test}-C_\text{train}$. We show the hardware-efficient ansatz with $N=4$ overparameterizes  for $M\ge M_\text{c}=4^N-1=255$ and $L_\text{c}=2^N=16$.
When the model is underparameterized $M\le M_\text{c}$, the training error in Fig.~\ref{fig:risk}a and test error in Fig.~\ref{fig:risk}b can be quite large. In contrast, the empirical generalization error in Fig.~\ref{fig:risk}c shows favorable scaling with $L$. However, note that the actual test error decreases in absolute value only slightly with $L$.
Overparameterization $M\ge M_\text{c}$ drastically reduces the training error to zero, and for $L\ge L_\text{c}$ allows us to find $C_\text{test}\approx0$.

\begin{figure*}[htbp]
	\centering	
\subfigimg[width=0.28\textwidth]{a}{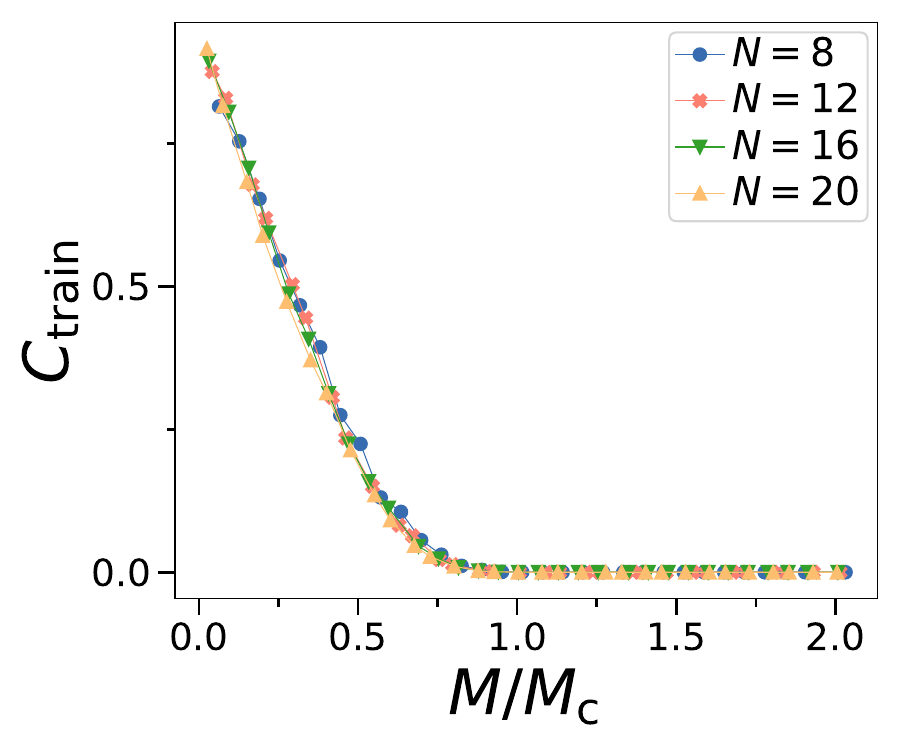}
\subfigimg[width=0.28\textwidth]{b}{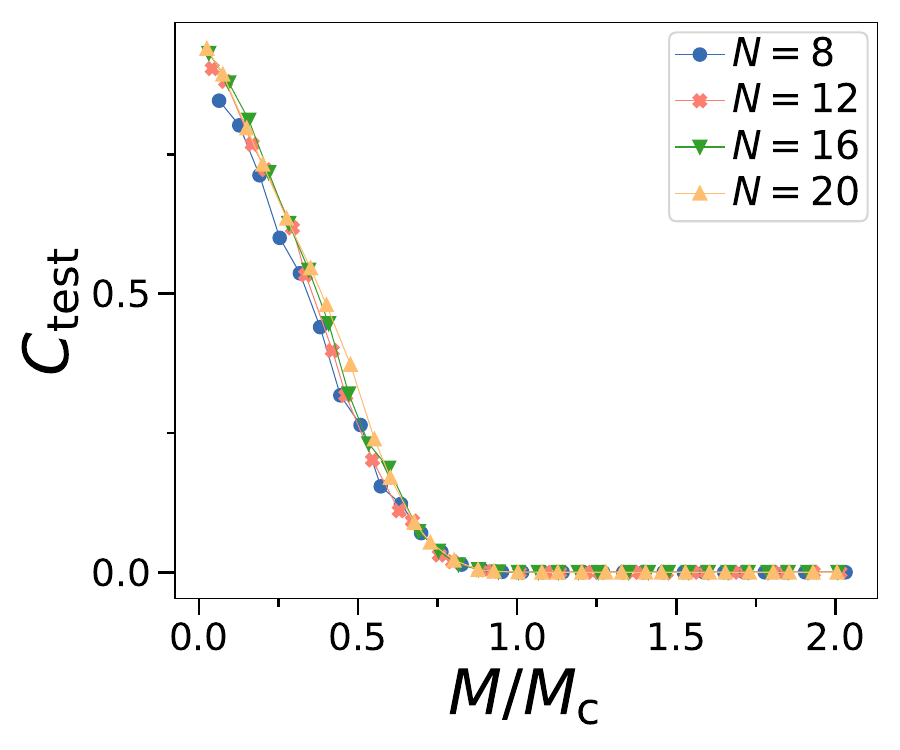}
\subfigimg[width=0.28\textwidth]{c}{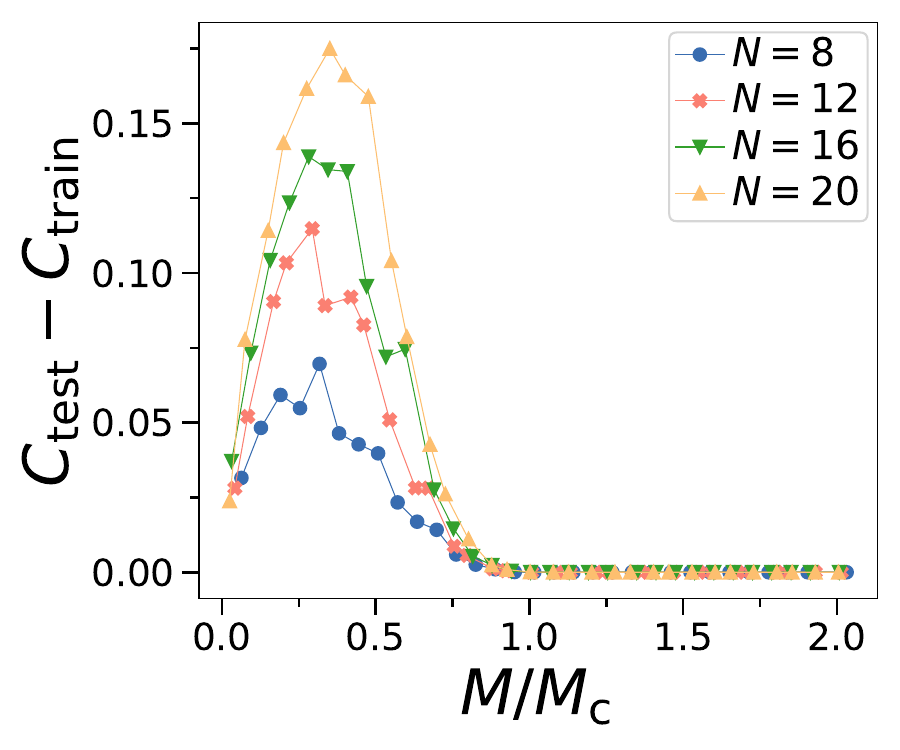}
	\caption{Learning with $U_{XY}$ and $\ket{\psi_\ell}\in W_{p=1}$ with fixed overcomplete data $L=40\gg L_\text{c}$ for $N=16$.
 \idg{a} $C_\text{train}$ against $M$ for different $N$, where $M_\text{c}=N^2-1=255$.
 \idg{b} $C_\text{test}$ against $M$ for different $N$. 
 \idg{c} Empirical generalization error $C_\text{test}-C_\text{train}$ against $M$.
	}
	\label{fig:singleRisk}
\end{figure*}

In Fig.~\ref{fig:singleRisk} we study  generalization and empirical generalization error for $U_{XY}$ ansatz and symmetric data $W_{p=1}$. In Fig.~\ref{fig:singleRisk}a,b we see that $C_\text{train}$ and $C_\text{test}$ decreases with $M$, and reaches near-zero for $M\ge M_\text{c}$.
In Fig.~\ref{fig:singleRisk}c we plot the empirical generalization error $\zeta=C_\text{test}-C_\text{train}$ against $M$. We find that $\zeta$ first increases, then decreases with $M$. We also note that $\zeta$ increases with $N$.

\end{document}